\documentclass[10pt]{article}
\usepackage{amssymb,amsfonts,amsmath}
\usepackage{graphicx}
\usepackage{cite}
\usepackage{color}

\usepackage[labelfont=bf,labelsep=period,justification=raggedright]{caption}

\makeatletter
\renewcommand{\@biblabel}[1]{\quad#1.}
\makeatother

\date{}

\usepackage{epsfig}
\usepackage{amsmath}
\usepackage{amsfonts}
\usepackage{amssymb}
\usepackage{graphicx}
\usepackage{subfigure}
\begin{document}
\begin{flushleft}
{\Large
\textbf{Coexistence in an inhomogeneous environment}
}
\\
Shlomit Weisman$^{1}$, David A. Kessler$^{1,\ast}$\\
\bf{1} Department of Physics, Bar-Ilan University, Ramat-Gan 52900 Israel
\\
$\ast$ E-mail: kessler@dave.ph.biu.ac.il
\end{flushleft}

\section*{Abstract}
We examine the two-dimensional extension of the model of Kessler and Sander of competition between two species identical except for
dispersion rates. In this class of models, the spatial inhomogeneity of reproduction rates gives rise to an implicit cost of dispersal, due to the tendency to leave favorable locations.  Then, as in the Hamilton-May model with its explicit dispersal cost, the tradeoff between dispersal case and the beneficial role of
dispersal in limiting fluctuations, leads to an advantage of one dispersal rate over another, and the eventual extinction of the disadvantaged species.
In two dimensions we find that  while the competition leads to the elimination of one species at high and low population density, at intermediate densities
the two species can coexist essentially indefinitely. This is a new phenomenon not present in either the one-dimensional form of the Kessler-Sander model nor in the totally connected Hamilton-May model, and points to the importance of geometry in the question of dispersal.

\section*{Introduction}

A central question in ecology is how species determine their migration rate. Migration entails a number of possible benefits to the population, among them reducing overcrowding areas
and the discovery of new resources to exploit. There are also significant costs associated
with migration. Especially in patchy environments, there may be a significant risk of death (due to predation or other exogenous causes) in
trying to cross over unsuitable areas to find a colonizable site. Presumably, migration
rates are under evolutionary control, and so can in principle be adjusted to some optimal
level. There have been a large number of studies in this field, both from an analytical
and a numerical perspective.\\
The current work is an extension of the work of Kessler and Sander~\cite{kesslersander} reconciling two seemingly contradictory approaches to this question of
optimal dispersal.  On the one hand, there has been extensive study of the rate equations approach to modeling. This approach had its beginnings in a  result of Hastings~\cite{Hastings}, who showed both in a two patch system and in a continuous one-dimensional system that the species with the slower dispersal rate is stable with respect to the introduction of a faster dispersing species, always driving the faster  species to extinction, as long as the two patches were not identical.    
 This is due to the fact that the
overall effect of dispersal was to take individuals from favorable areas to less suitable
places, leaving the slow species in control of the favorable site.  The higher production rate of progeny at  the favorable site eventually leads to a takeover of the poorer site as well.  Thus,  dispersal in a inhomogeneous environment carries with it an implicit cost. This finding was
reinforced by Dockery and coworkers \cite{Dockery}. Examining the continuous one-dimensional version of the Hastings model,
 they studied the reaction-diffusion system:
\begin{equation}
\begin{array}{ll}
\dot{f}=D_f{f}''+\alpha(x)f-\beta f(f+s)\\
\dot{s}=D_s{s}''+\alpha(x)s-\beta s(f+s)
\end{array}
\label{eq:reaction diffution eq Dockery}
\end{equation}
Here, $f$ and $s$ represent the local population density of the fast ($f$) and slow ($s$) migrating
species, with $D_{f} > D_{s}$ . The inhomogeneous distribution of resources is modeled by a
spatially varying growth rate, $\alpha(x)$, common to the two species. The competition is of
logistic type, with each species competing equally against both fellow members of the
same species and members of the other variety. The Hastings result is that the pure slow state  with $f = 0$ is linearly stable, and the pure fast state, $s=0$ is unstable.  Dockery, et al. were able to prove the stronger result that, independent of the
exact form of $\alpha(x)$ and of the initial conditions, the pure slow state is the only global
attractor and the faster species always went to extinction. The implication of both these works is that any mutation resulting
in a lower level of migration should fix in the population, driving the migration rate
eventually to zero.

This strong result is in sharp contrast to what occurs in the individual based models originated by Hamilton and May~\cite{hm}.  Looking at a system with an
infinite site of patches fully connected by dispersal, and with an explicit cost of dispersal, they showed that the evolutionary stable stategy was a large level of dispersal. This optimal level of dispersal approached unity as the dispersal cost was lowered to zero, and surprisingly, remained finite even when the dispersal cost approached 100\% of the dispersers. 

Kessler and Sander~\cite{kesslersander} reconciled these results by showing that the demographic fluctuations inherent in the individual based model, and absent in the rate equation approach, were responsible for the very different results of the two models.  By adding a noise term to the differential equation, Eq. (\ref{eq:reaction diffution eq Dockery}), they showed that fluctuations have a negative effect and that dispersal, by reducing such fluctuations by uniformizing the population in space, has a beneficial effect.  Thus, the tradeoff between the dispersal cost and the fluctuation reduction benefit gives rises to a nonzero optimal value of dispersal rate. This result underlies the decrease of the optimal dispersal rate with increasing population density seen in the Hamilton-May-Comins~\cite{hm1} extension of their model to arbitrary population density, since increasing population density decreases fluctuations and so reduces the beneficial aspects of dispersal.

To verify their findings, Kessler and Sander studied an individual based version of the one-dimensional Dockery model.
The system has two species of individuals, fast
and slow, residing on sites of a one-dimensional lattice $x_i=0,1,\ldots L$ with no-flux boundary conditions. The system is designed to reproduce a spatially discretized version of the Dockery model in the deterministic limit. All individuals have a probability of $\alpha(x_{i})dt$ of reproducing in some small time interval $dt$, depending on their location $x_{i}$. Furthermore, individuals at $x_{i}$ have a concentration-dependent probability of death, $(f_{i} + s_{i})\beta dt$, where $f_{i}$ and $s_{i}$ are the numbers of fast and slow individuals sharing the particular site. They also move to a nearest-neighbor site with probability  $D_{f,s}dt$, depending on their identity. Kessler and Sander chose to examine a distribution of resources corresponding to two
``oases" of food separated by a desert:
\begin{equation}
\alpha(x)=1+\eta \cos \left(\dfrac{2\pi(x+0.5)}{L}\right)
\label{eq:surface D1}
\end{equation}
They set $\beta=\dfrac{1}{N_{0}}$, so that in the absence of diffusion, the total population at each site would saturate at the local carrying capacity $N_{0}\alpha(x_{i})$. The results of this model showed that, for a given level of inhomogeneity $\eta$, when $N_0$ was large, the rate equations correctly predict that the slower species wins. On the other hand, for $N_0$ small, the ability of the faster species to use diffusion to minimize fluctuations led to its victory, against the predictions of the rate equations. One surprise was that the boundary between large and small $N_0$ was surprisingly large, in the range of hundreds per lattice site, against the naive expectation that the rate equations would be reliable for $N_0$'s of order 10 or so.  The interplay between inhomogeneity and $N_0$ was consistent with the observation that in the Hamilton-May-Comins~\cite{hm1} extension of the model, wherein an arbitrary carrying capacity is allowed, the evolutionary stable dispersal rate decreases with $N_0$, and with increasing dispersal cost.

\begin{figure}[h]
\begin{center}
\scalebox{0.9} {\includegraphics[width=\columnwidth]{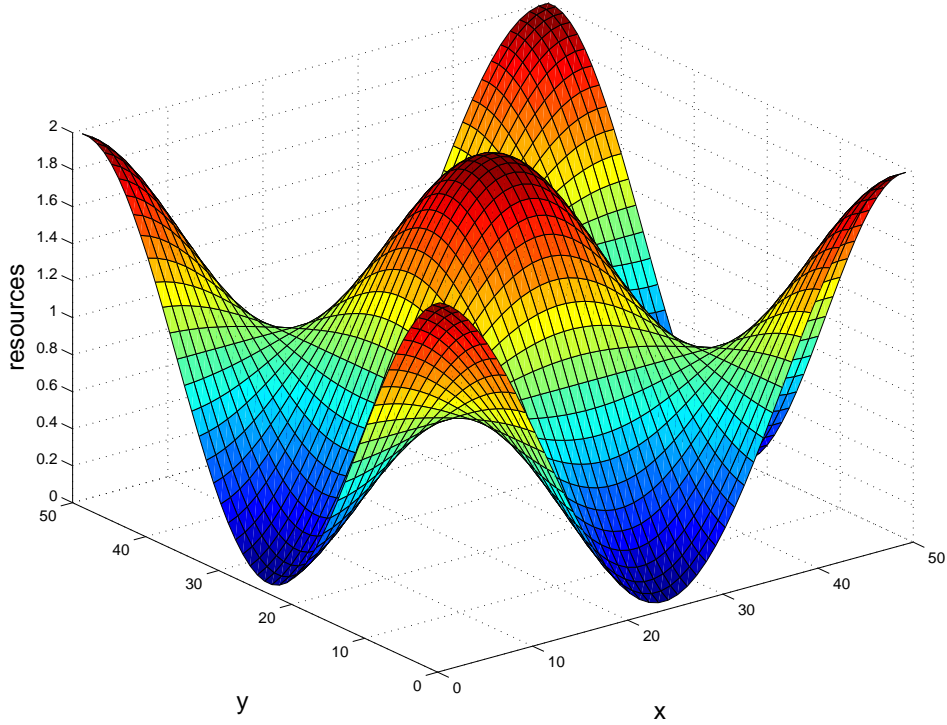}}
\end{center}
\caption{Plot of the local birthrate $\alpha(x,y)$ in our $50\times 50$ system.}
\label{fig:resources 2D}
\end{figure}

The model we examine herein is the natural two-dimensional generalization of the Kessler-Sander model. The system consists of a square two-dimensional lattice, $x_i,y_i = 0,1,\ldots L$. The dynamics is exactly the same as in the one dimensional case, with the dispersal move now occurring to one of the four nearest-neighbor sites. We choose for the inhomogeneity a ``checkerboard" pattern
\begin{equation}
\alpha(x,y)=1+\eta \cos \left(\dfrac{2\pi(x+0.5)}{L}\right)\cos \left(\dfrac{2\pi(y+0.5)}{L}\right)
\label{eq:surface D2}
\end{equation}
with oases in the corners and the center and deserts in-between, see Fig. \ref{fig:resources 2D}. In all our simulations, we have set $L=50$.

\section*{Results}

\begin{figure}
  \centering
  a{\label{fig:f1a}\includegraphics[width=0.31\textwidth]{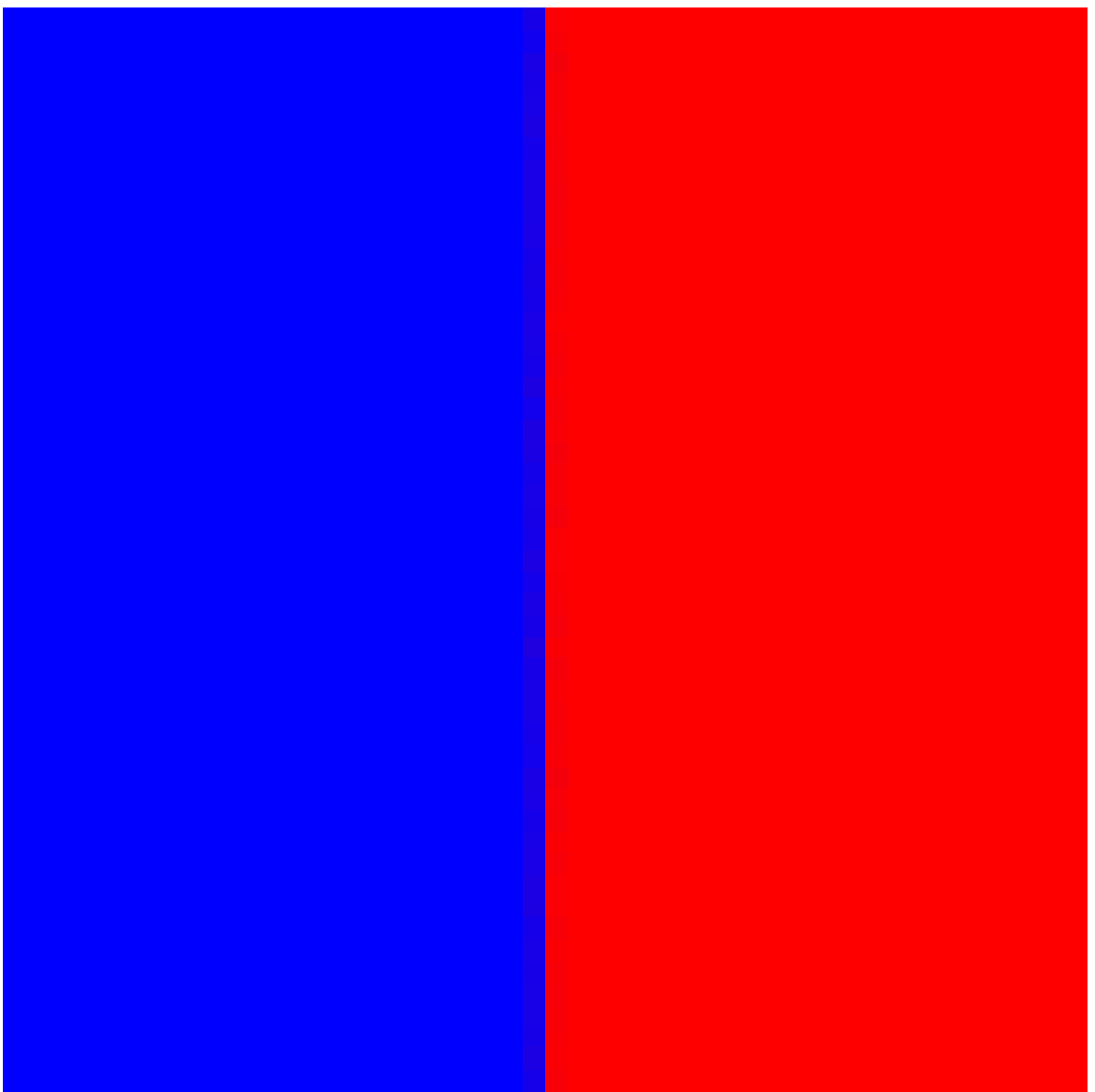}} 
  b{\label{fig:f1b}\includegraphics[width=0.31\textwidth]{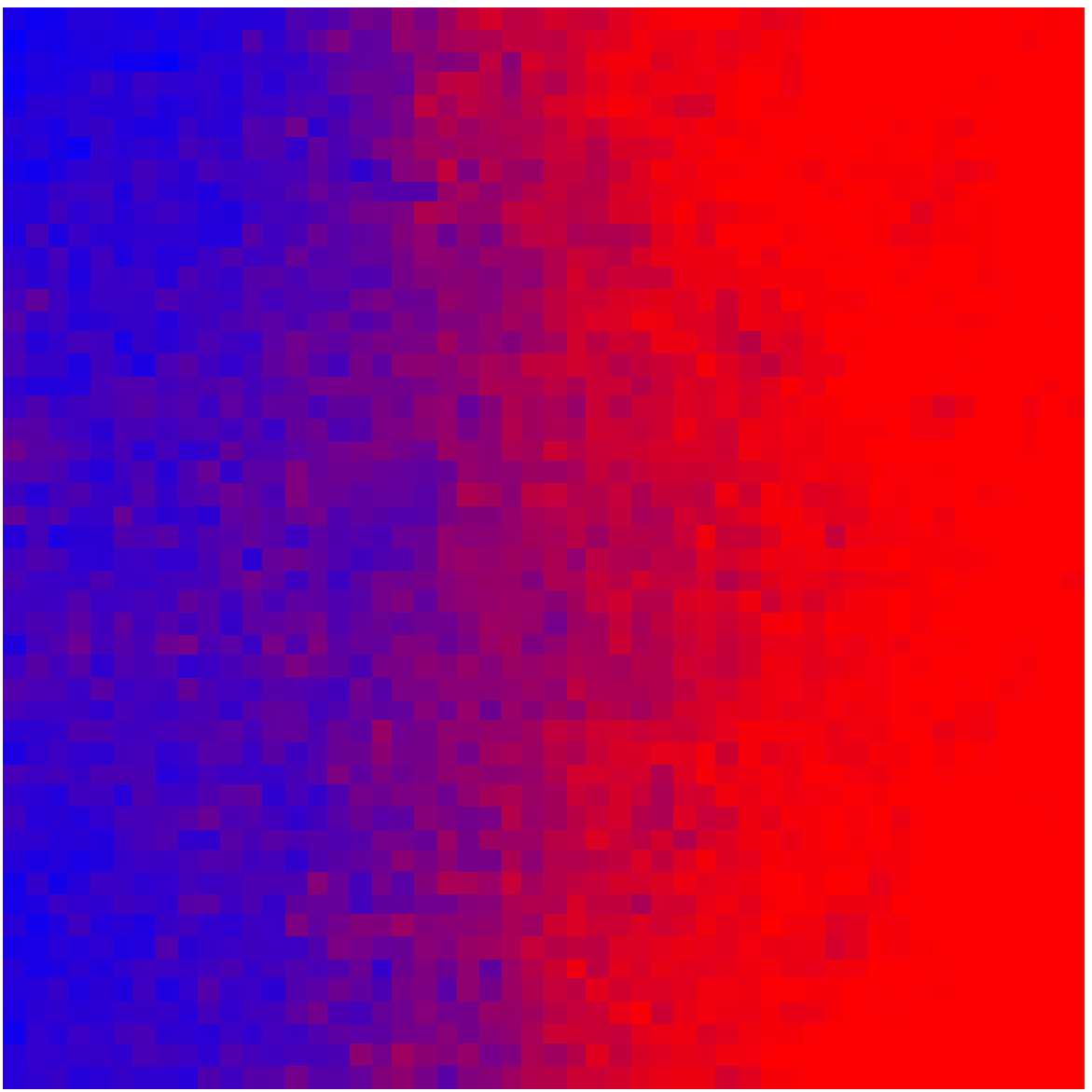}}
  c{\label{fig:f1c}\includegraphics[width=0.31\textwidth]{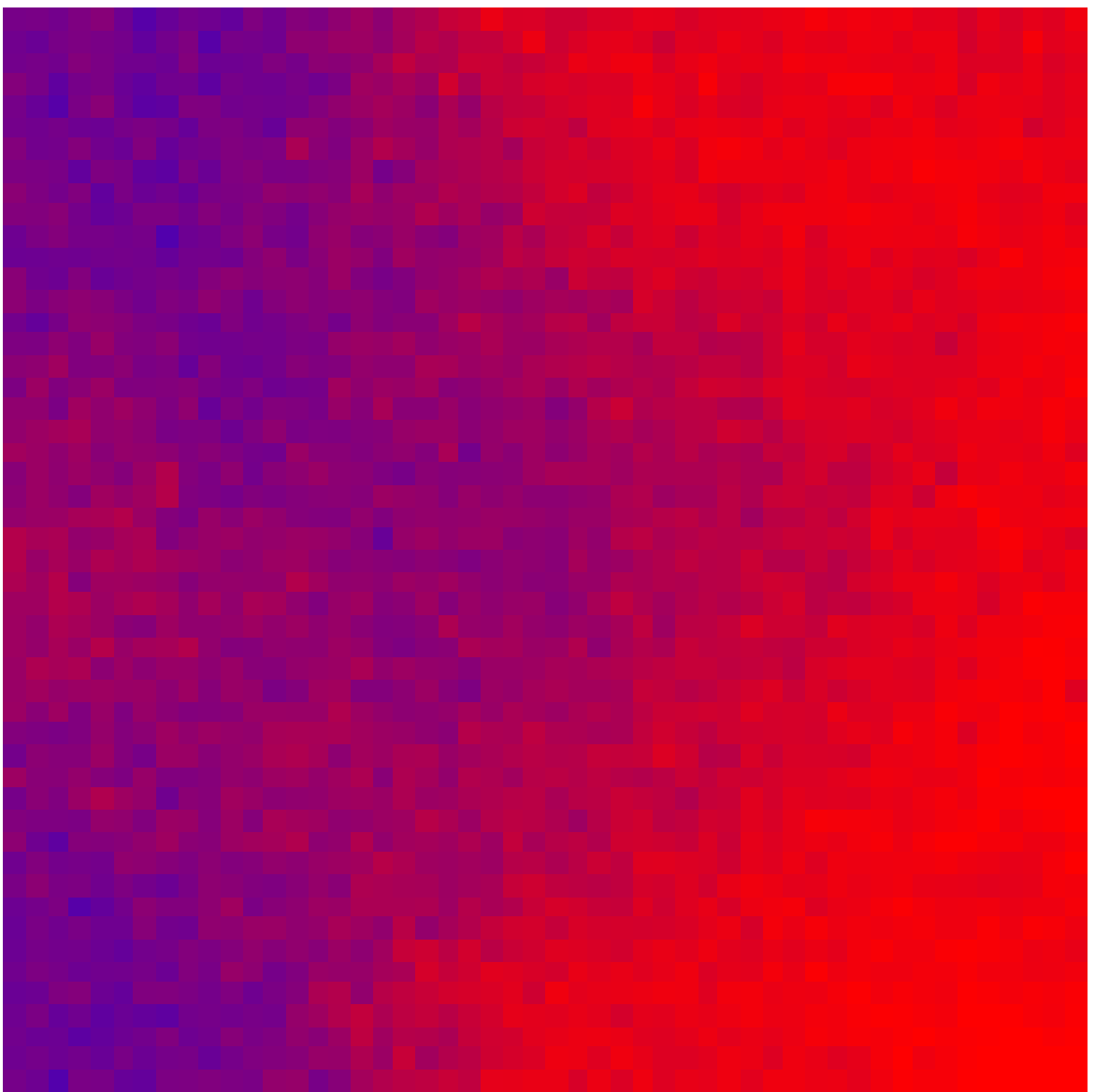}}\\
\centering
  d{\label{fig:f1d}\includegraphics[width=0.31\textwidth]{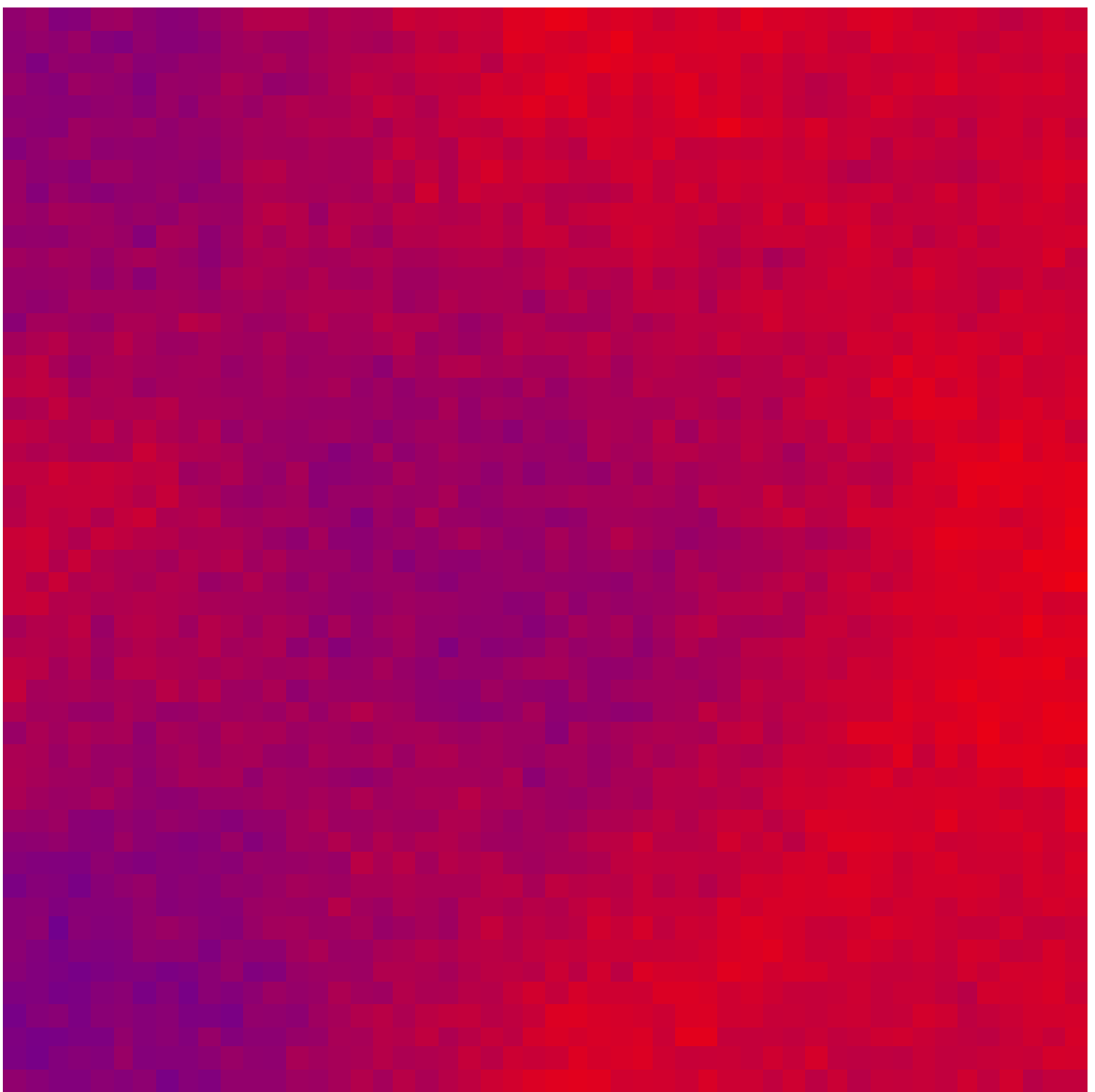}} 
  e{\label{fig:f1e}\includegraphics[width=0.31\textwidth]{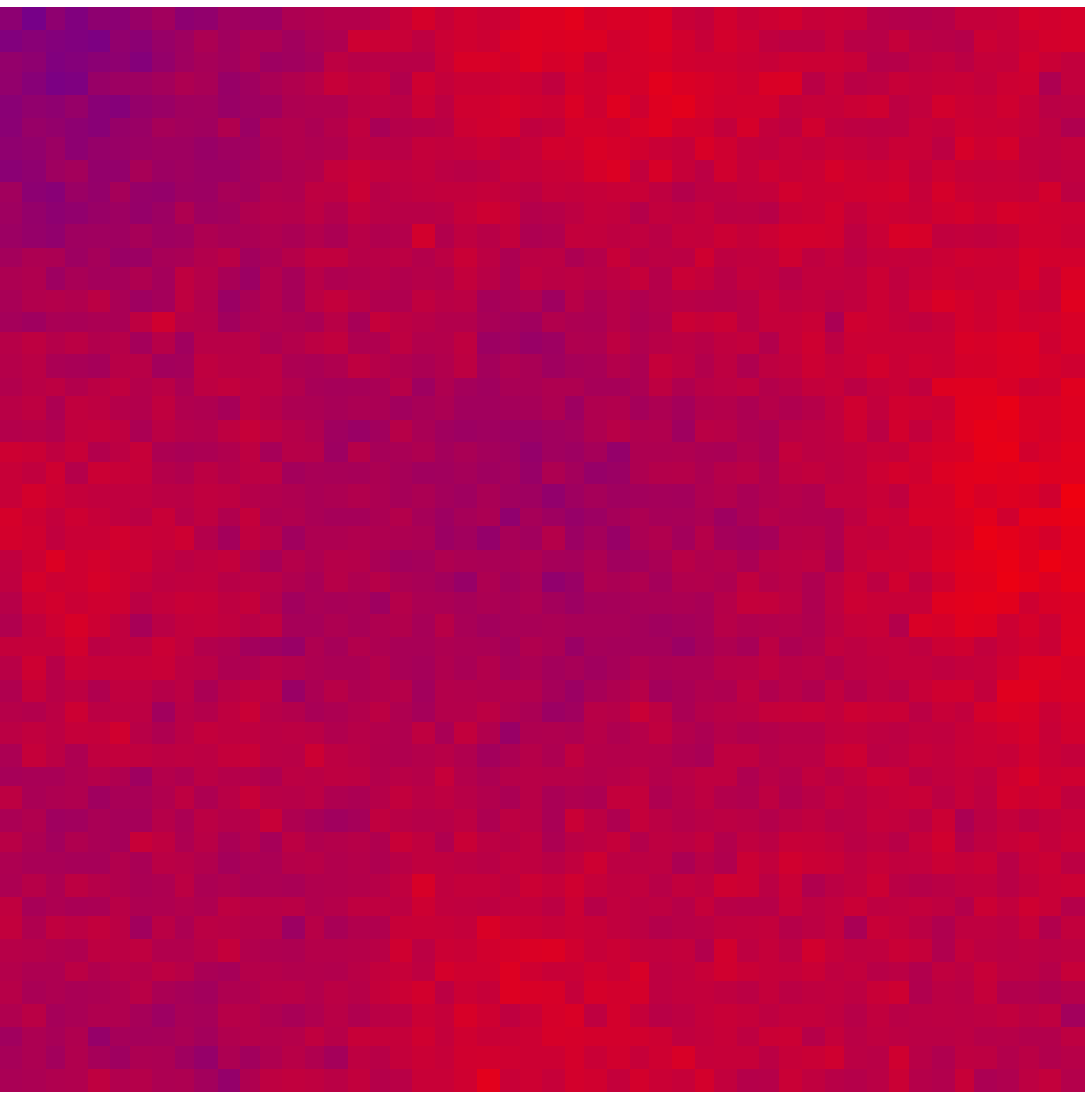}}
  f{\label{fig:f1f}\includegraphics[width=0.31\textwidth]{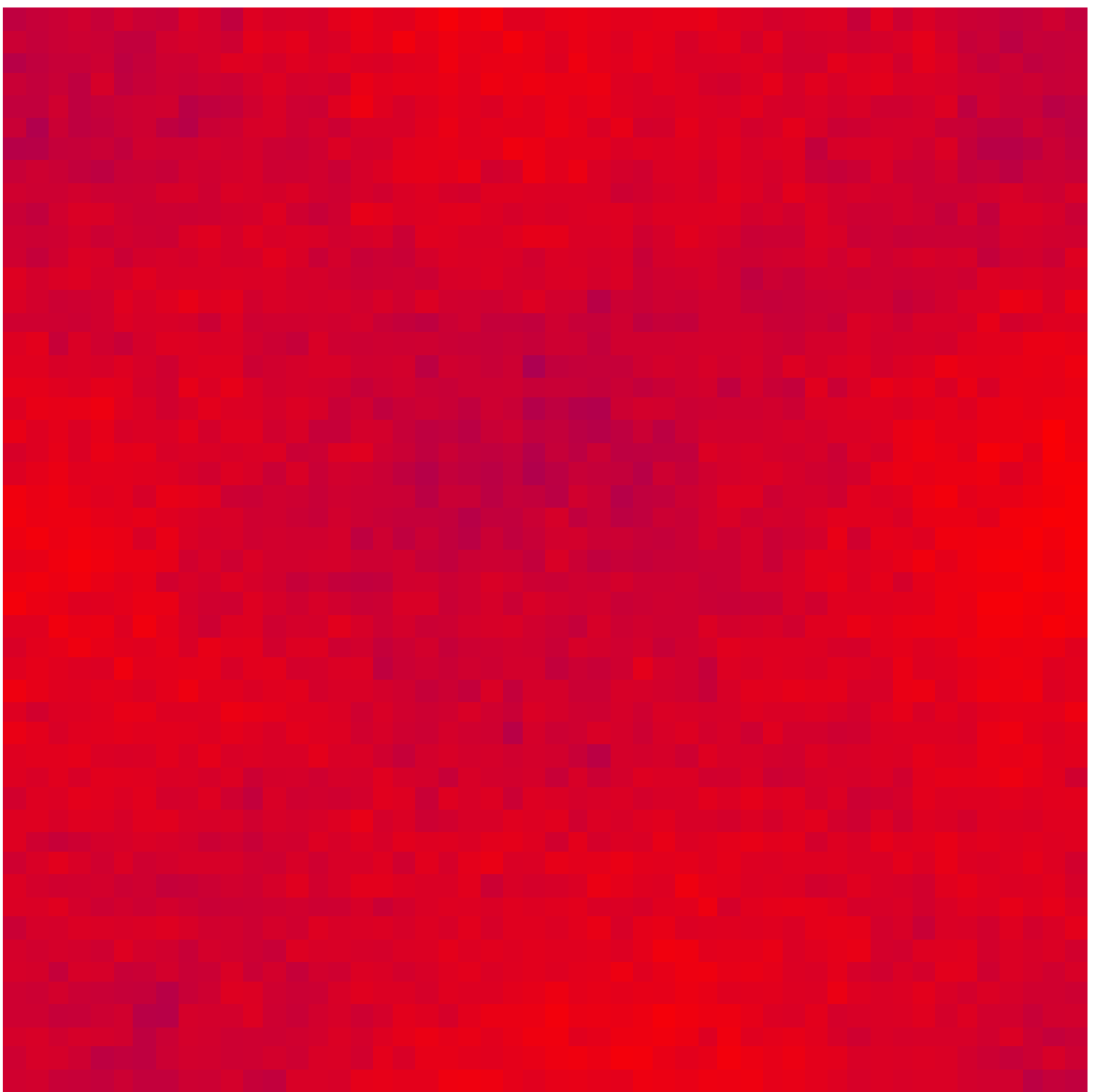}}
\caption{Snapshots from a simulation with the parameters $\eta=0.4, N_{0}=400$. (a) The initial state of the system. (b)-(e) The fasts initially fare better in the ``desert'' and the slows in the ``oases''. (b) the system after 100 steps, (c)  450 steps, (d)  1000 steps, (e) 1300 steps. (f) Essentially total victory of the fasts after 3000 steps.}
  \label{fig:f_win}
\end{figure}

\begin{figure}
  \centering
  a{\label{fig:sa}\includegraphics[width=0.31\textwidth]{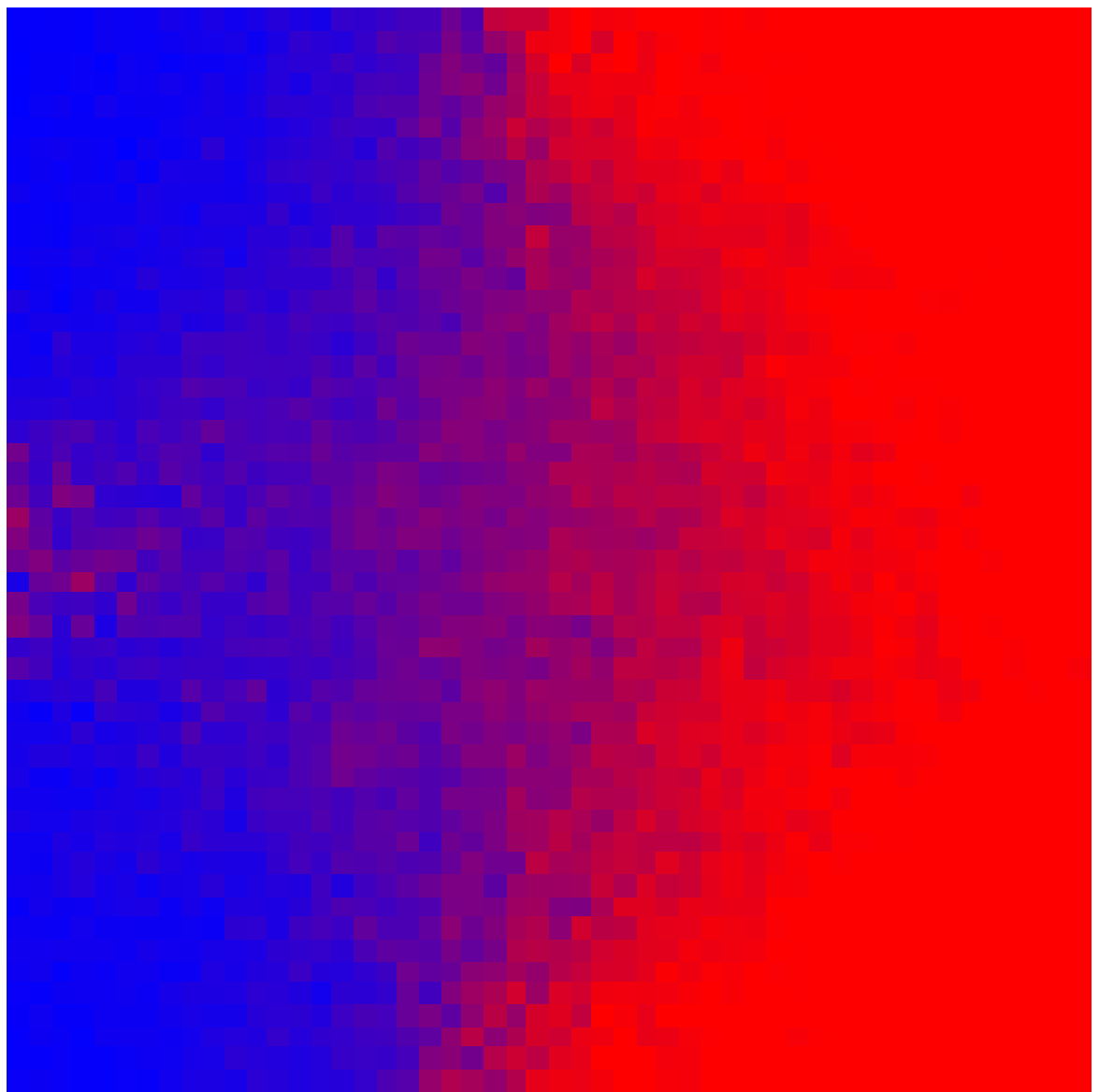}}
  b{\label{fig:sb}\includegraphics[width=0.31\textwidth]{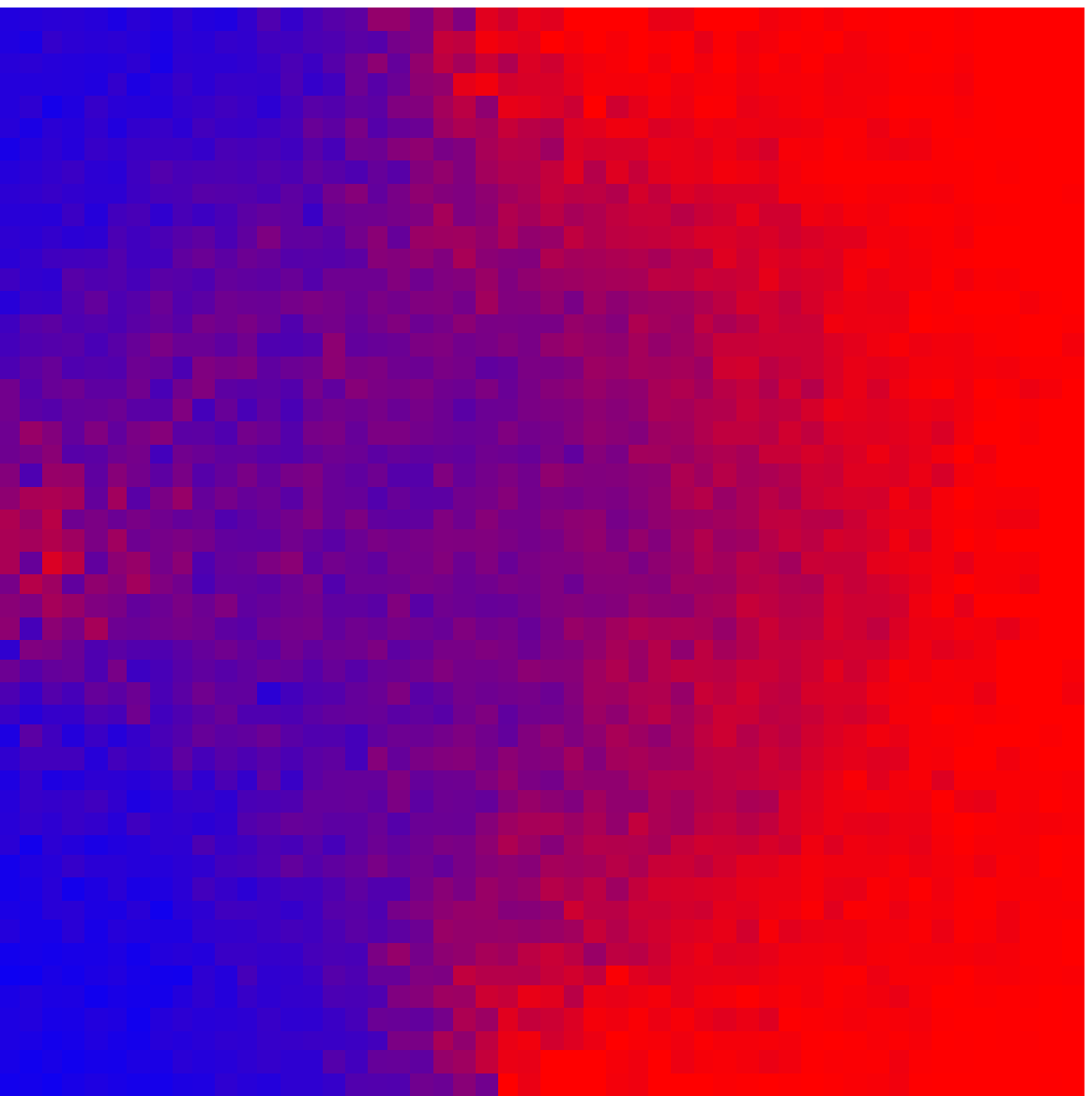}}\\
\centering
  c{\label{fig:sc}\includegraphics[width=0.31\textwidth]{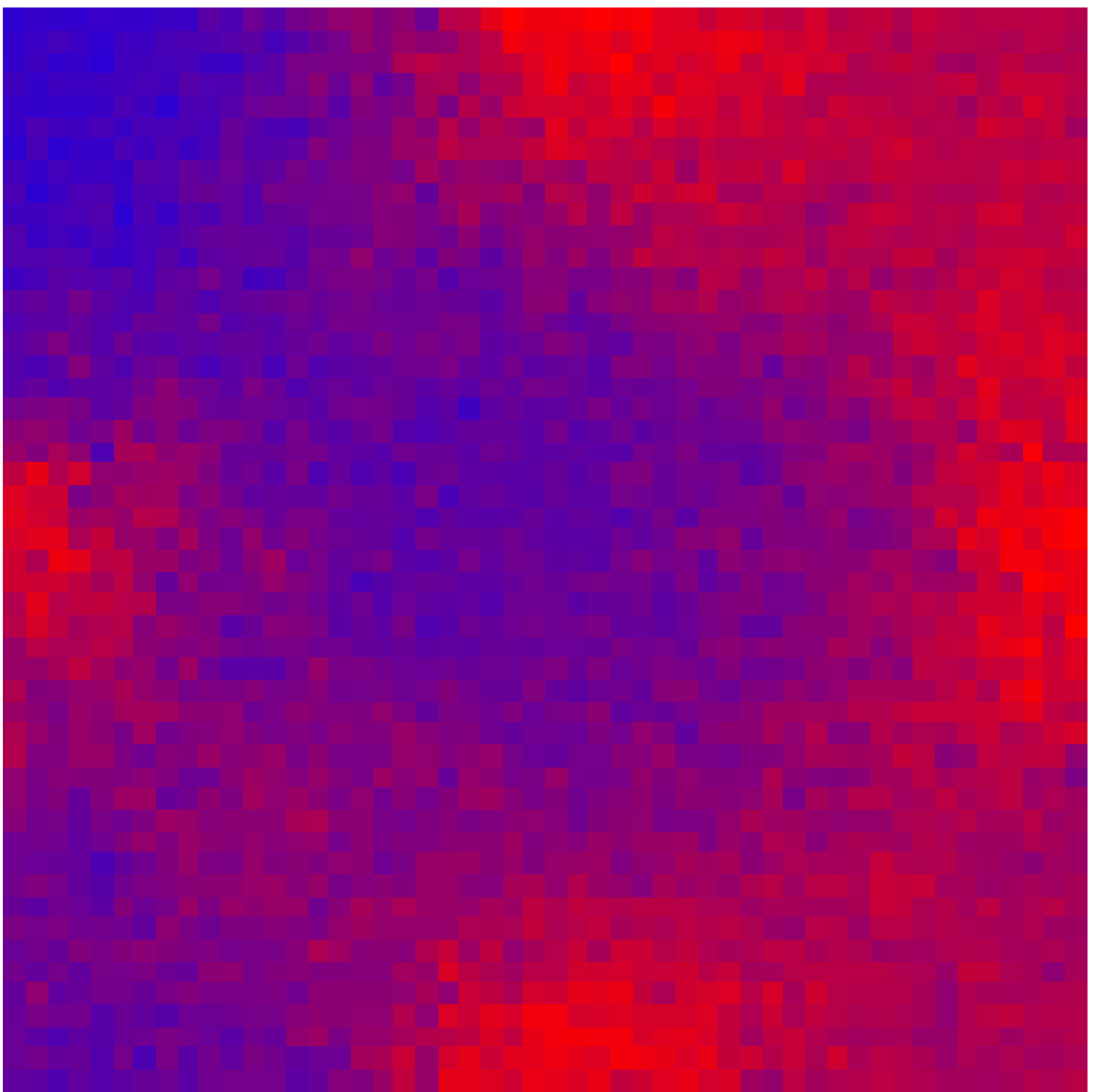}} 
  d{\label{fig:sd}\includegraphics[width=0.31\textwidth]{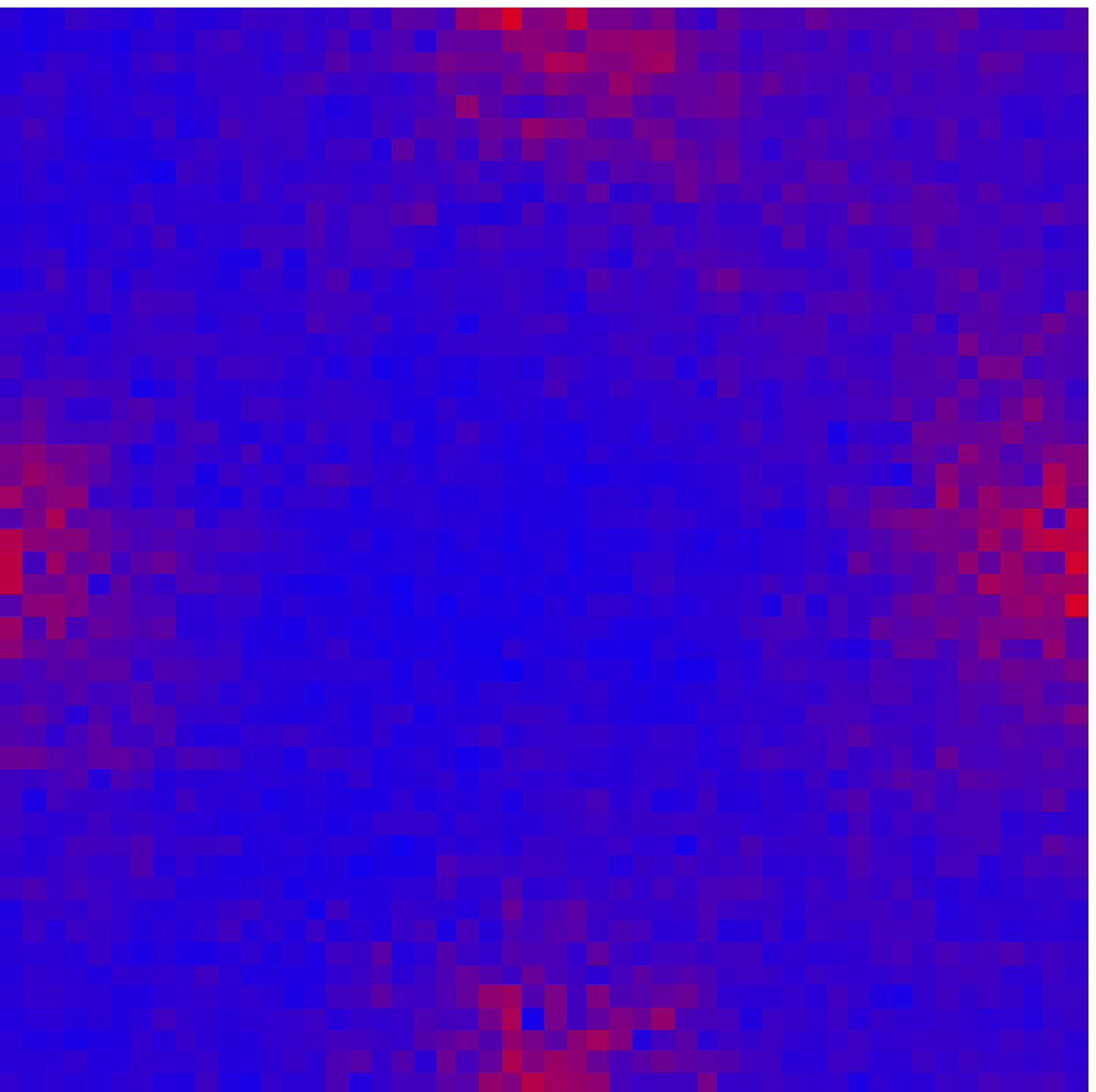}}
\caption{Snapshots from a simulation with the parameters $\eta=0.4, N_{0}=700$. (a) The system after 100 steps, (b)  200 steps, (c)  1000 steps, (f) almost complete victory of the slows after 4500 steps.}
  \label{fig:s_win}
\end{figure}

\begin{figure}
  \centering
  a{\label{fig:coa}\includegraphics[width=0.31\textwidth]{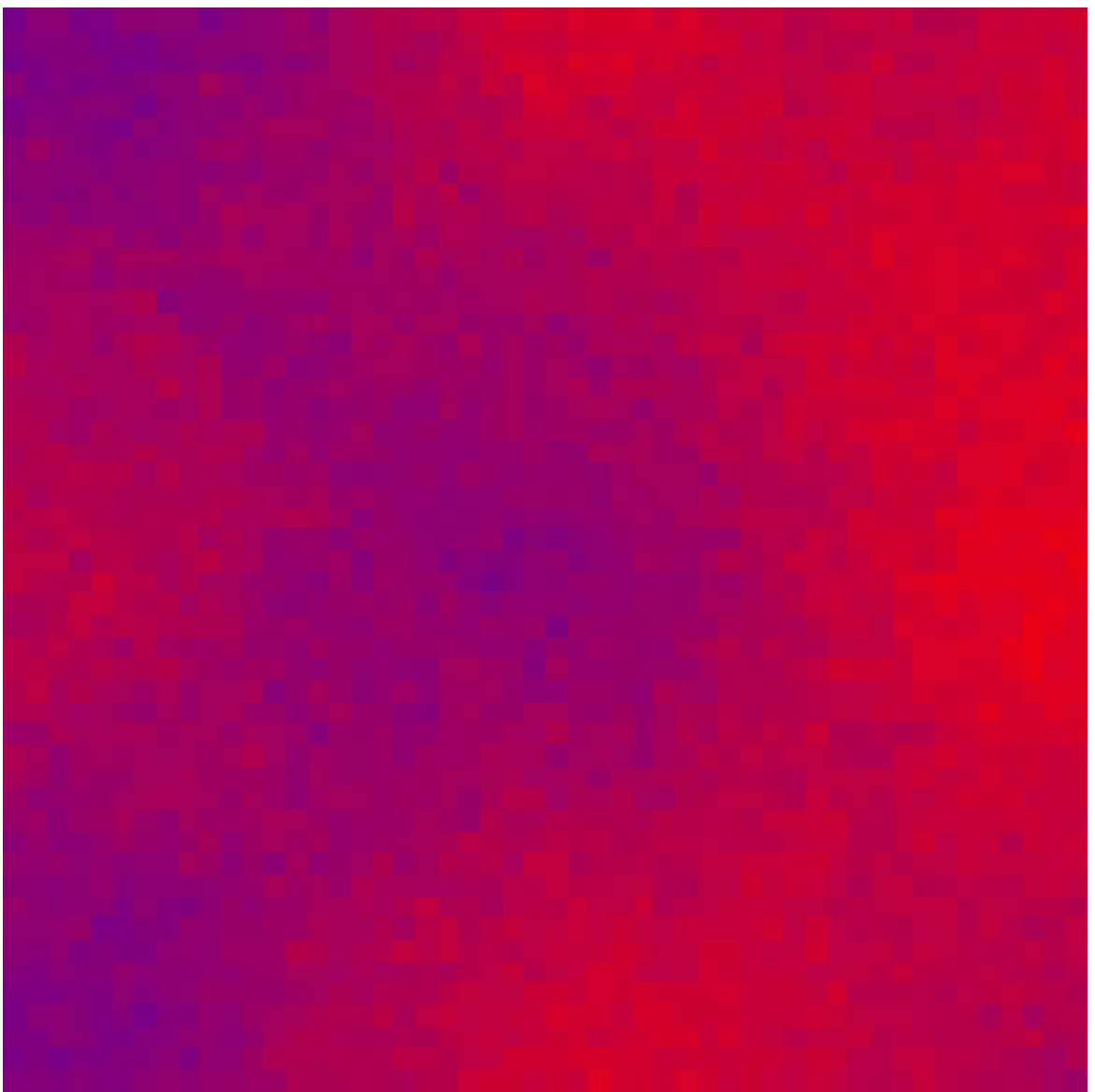}}
  b{\label{fig:cob}\includegraphics[width=0.31\textwidth]{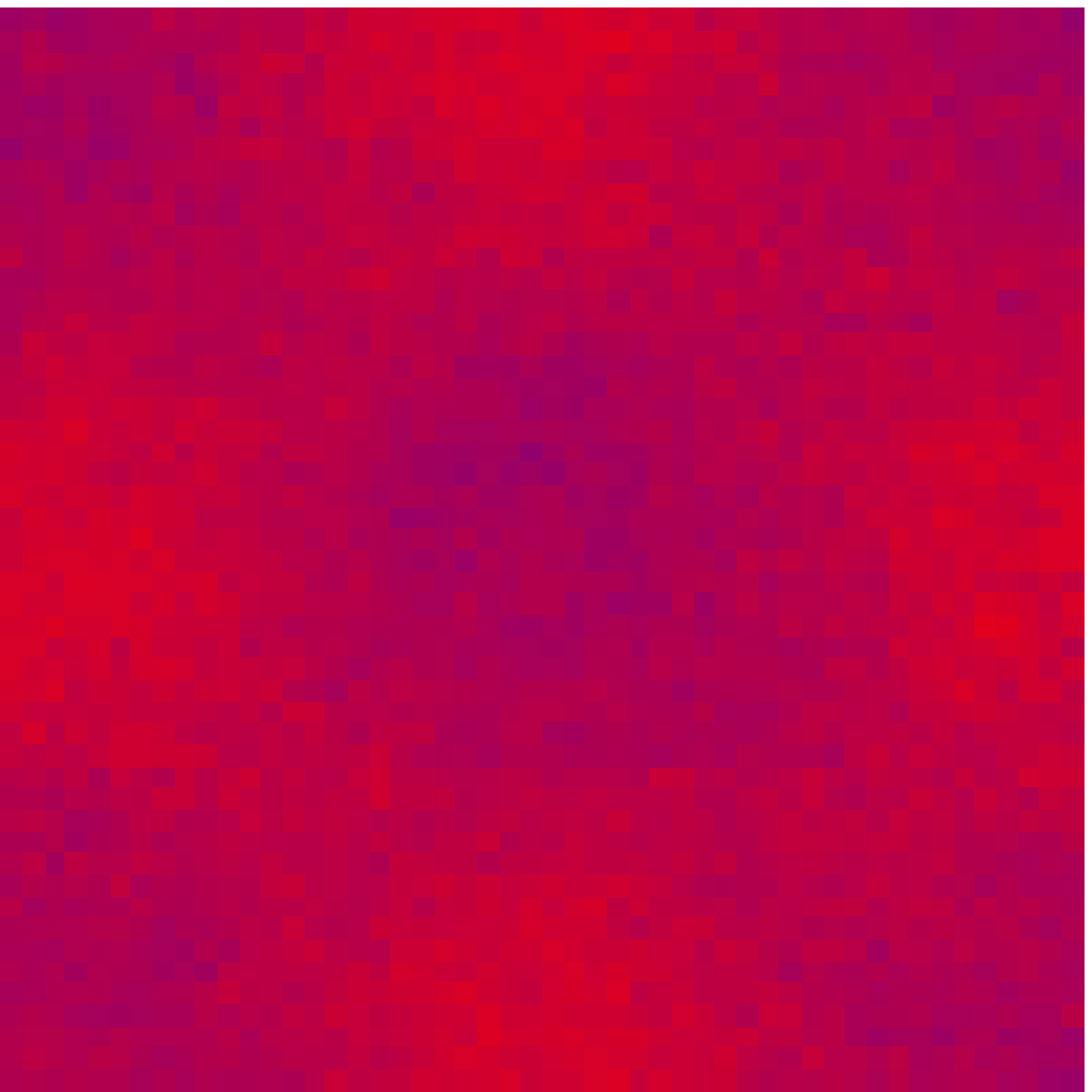}}\\
\centering
  c{\label{fig:coc}\includegraphics[width=0.31\textwidth]{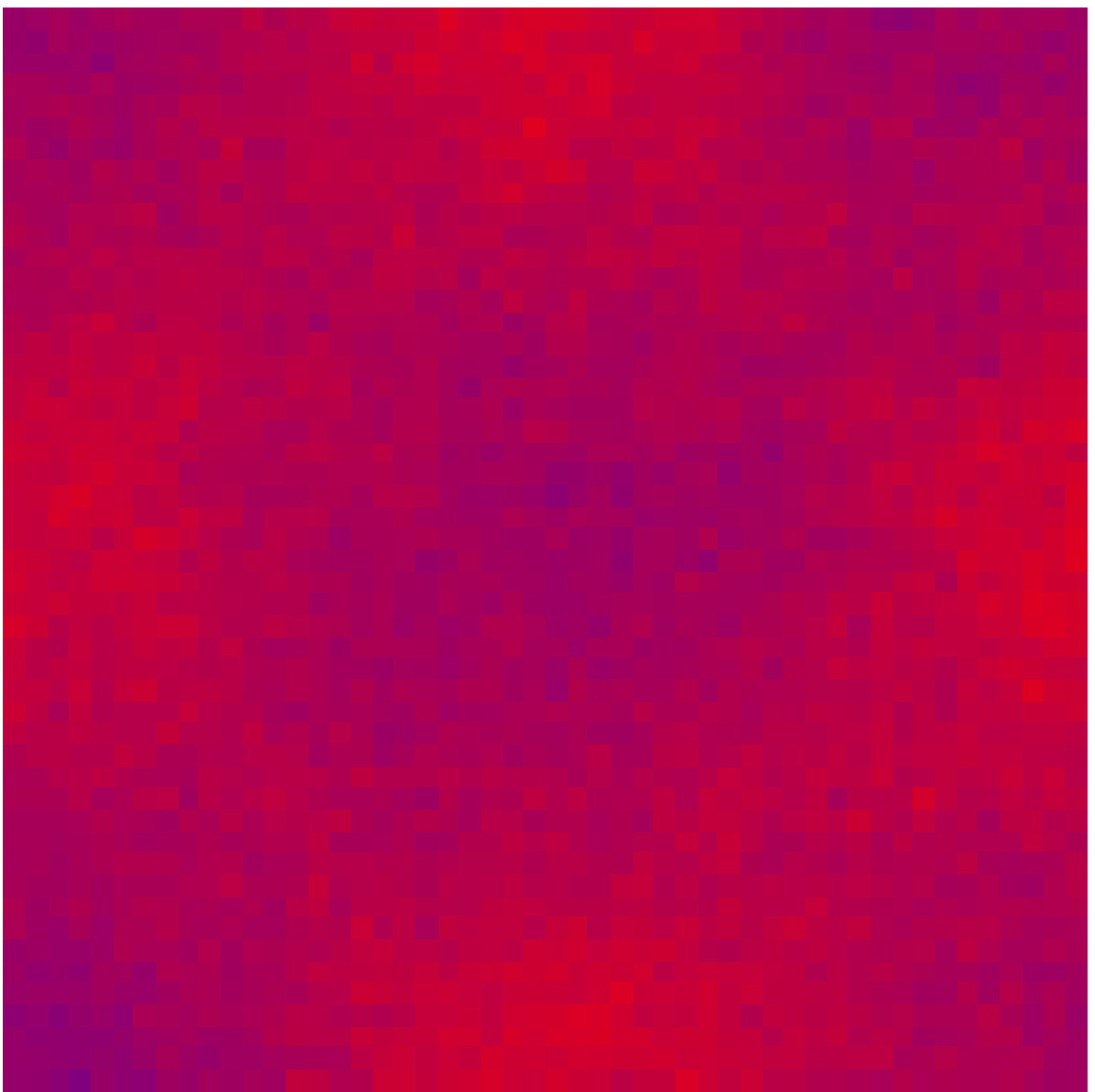}} 
  d{\label{fig:cod}\includegraphics[width=0.31\textwidth]{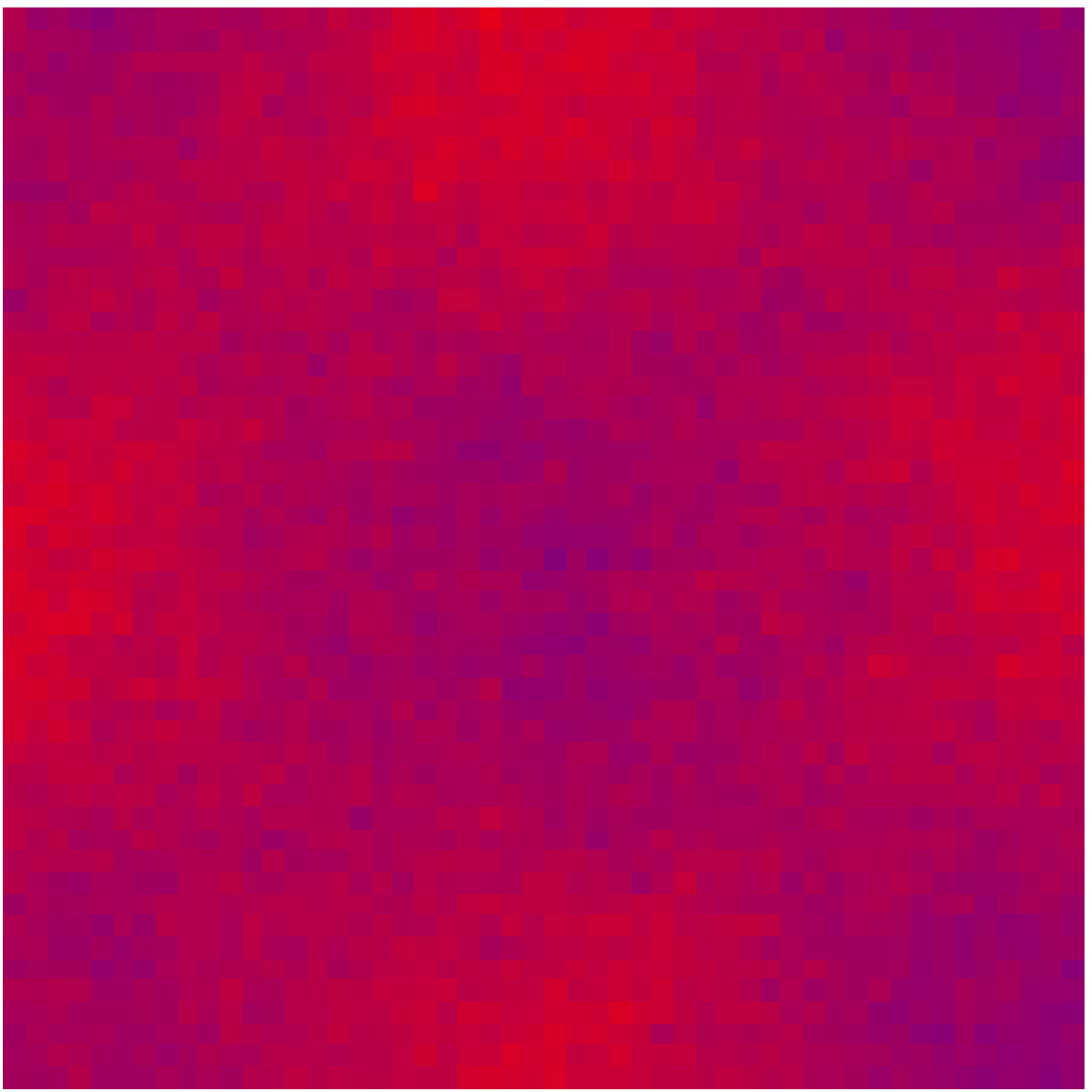}}
\caption{Graphic simulation with the parameters $\eta=0.4, N_{0}=600$. (a) the system after 1000 steps, (b) the system after 5800 steps, (c) the system after 11800 steps, (f) the system after 15000 steps, there are no essential changes during the time.}
  \label{fig:coexistence}
\end{figure}

Figs. 
\ref{fig:f_win}-\ref{fig:coexistence} display some snapshots from the two dimensional simulation. The color represents the relative abundance of fast and slow on a site, with all fast being pure red and all slow being pure blue. 
Fig.\ref{fig:f_win} displays a system with the parameters: $\eta=0.4, N_{0}=400$. Fig. \ref{fig:f_win}(a) displays the initial state of the system for all the runs, namely only slows in the left half of the system and only fasts in the right half. In Fig.\ref{fig:f_win}(b)-Fig.\ref{fig:f_win}(e) it is possible to see that initially the fasts predominate in the deserts and the slows in the oases.  Eventually, for this set of parameters, with its relatively low population density, the fast dispersing species encroach on the oases and drive the slows toward extinction, as seen in n Fig.\ref{fig:f_win}(f). 

Fig. \ref{fig:s_win} displays a system with the same degree of inhomogeneity, $\eta=0.4$ but a larger population density $N_{0}=700$. The system starts as in  Fig.\ref{fig:f_win}(a). Again, the fasts fare better in the deserts, but the eventual advantage accrues to the slows, as expected due to the higher population density. By Fig.\ref{fig:f_win}(d), after 4500 steps, the victory of the slows is essentially complete, with the slows in the ``deserts" making a valiant but futile last stand, in accord with the statement that the slows win at sufficiently large $N_0$.

The above results are consistent with the picture that emerged from Kessler and Sander's simulations of the one-dimensional model.  However, we show in 
Fig. \ref{fig:coexistence} snapshots of a system with the same inhomogeneity $\eta=0.4$ and an intermediate value of the population density, $N_{0}=600$. Again, the system starts as in Fig.\ref{fig:f_win}(a). In this figure it is clearly seen that despite the much longer run times compared with the above two cases, there is  essentially no change in the ratio between the fasts and slows after 15000 time steps. This behavior leads to the speculation that the state displayed in this figure will be preserved also at infinite time, with the system having settled into a state of stable coexistence.

\begin{figure}[h]
\begin{center}
 {\includegraphics[width=0.8\columnwidth]{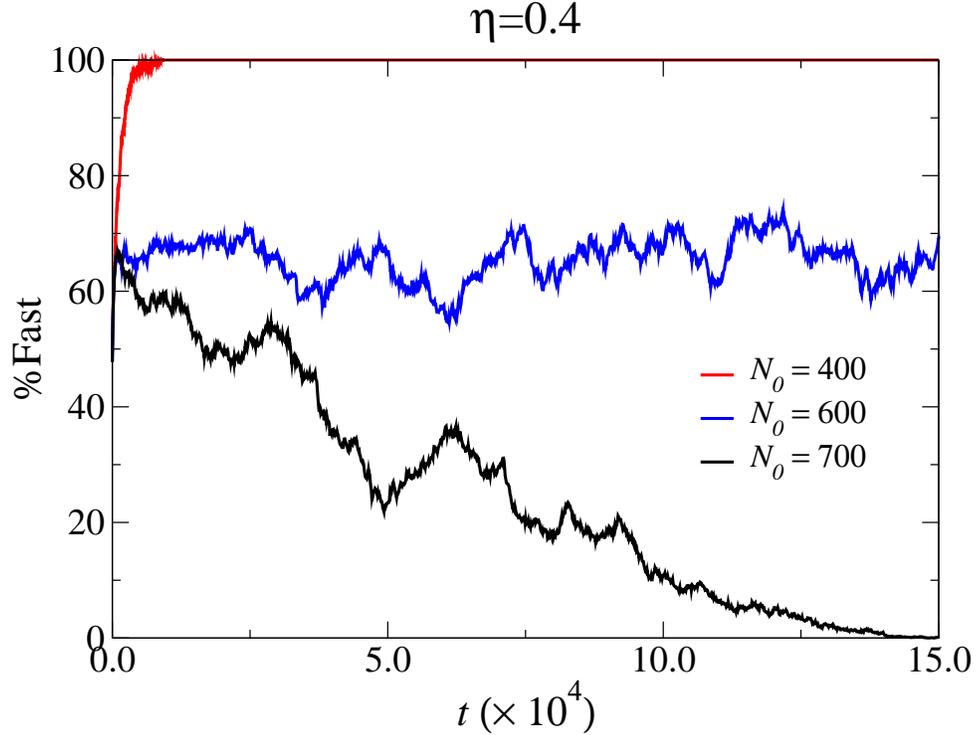}}
\end{center}
\caption{Percentage of fast individuals as a function of time for the simulations presented in Figs. \ref{fig:f_win}, \ref{fig:s_win} and \ref{fig:coexistence} with $\eta=0.4$ and $N_0=400,\, 600,\, 700$, respectively.}
\label{fig:perc_f_chess}
\end{figure}

Fig.\ref{fig:perc_f_chess} display the percentage of fast individuals in the entire system as a function of time for the three cases presented in the snapshots. For $\eta=0.4$, $N_0=400$
the percentage of fasts rises quickly to 100\%. While $\eta=0.4$, $N_0=700$ the percentage of fasts  decreases to zero.  It should be remarked that in this case the victory of the slows in clear from the spatial snapshot long before the total number of fasts has dropped significantly. For 
$\eta=0.4$, $N_0=600$ the  percentage of fasts fluctuates around 67.2\%, with no discernible secular trend.

\begin{figure}[h]
\begin{center}
 {\includegraphics[width=0.8\columnwidth]{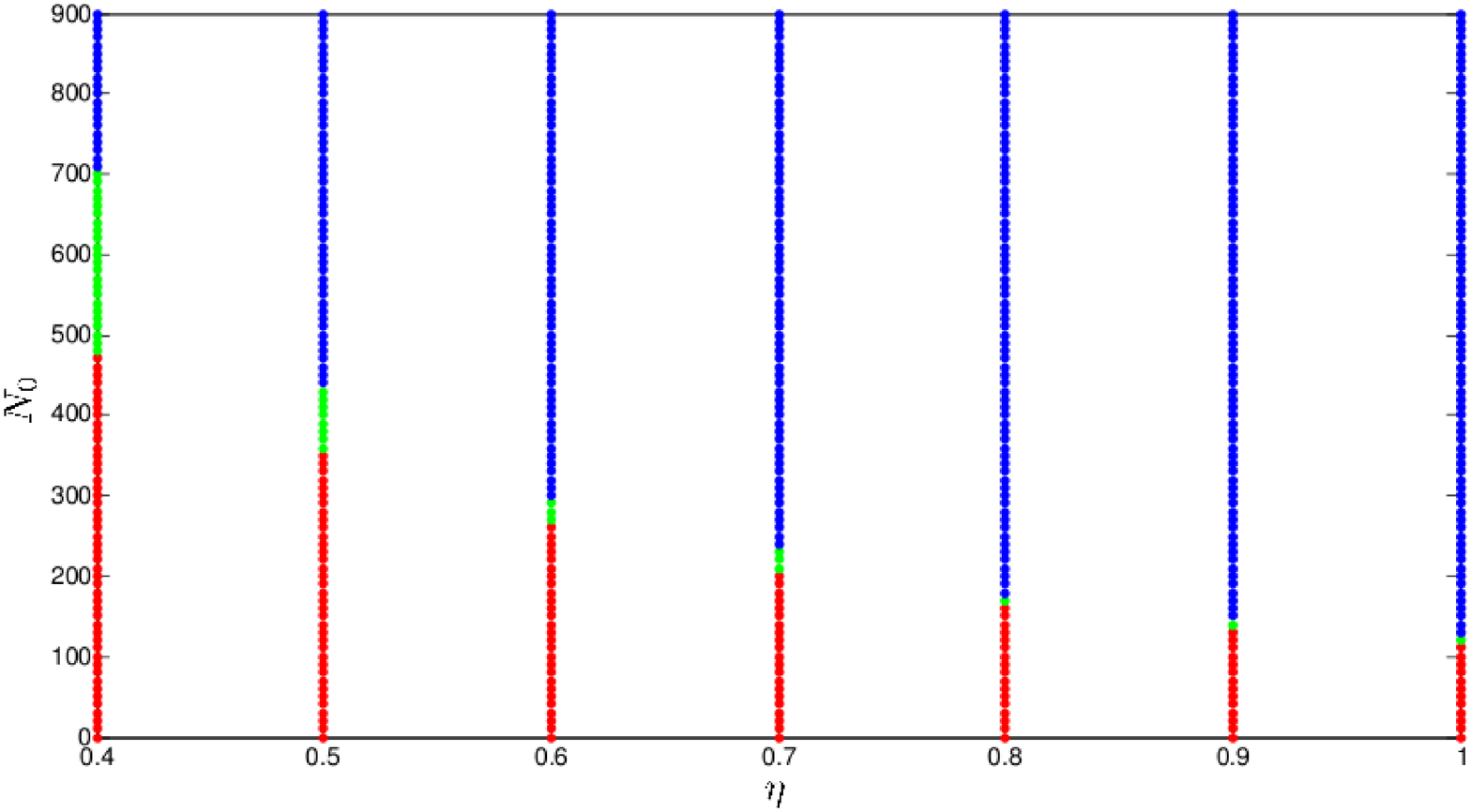}}
\end{center}
\caption{Results of the two dimension simulation. Red: fasts win, blue: slows win, green: coexistence}
\label{fig:points}
\end{figure}

To investigate further, we scan the $\eta$, $N_0$ parameter space, the results of which are summarized in
Fig.\ref{fig:points}. The blue color represents parameters in which the slows win, the red color represents parameters in which the fasts win. The green color  represents parameters with apparent coexistence. The overall trend of the dependence on $\eta$ and $N_0$ are as in the one dimensional model 
with higher inhomogeneity and higher $N_0$ both favoring the slows. The new feature is of course the intermediate coexistence phase, which is seen to shrink in extent as $\eta$ increases, at least for $\eta\ge 0.4$.

\begin{figure}
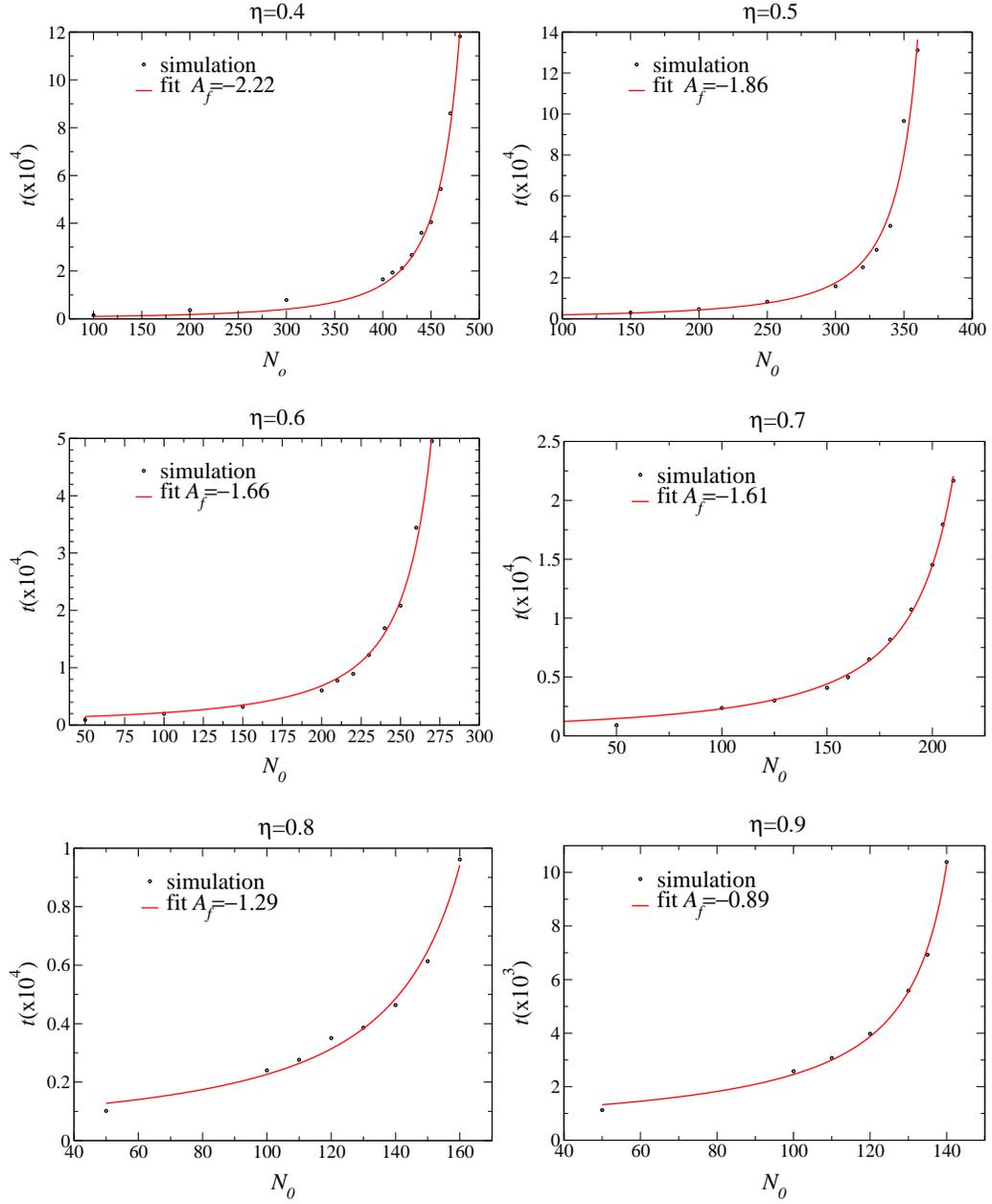

 \centering
  {\label{fig:0.4u}\includegraphics[width=0.42\textwidth]{time_0p4u.eps}}
  {\label{fig:0.5u}\includegraphics[width=0.42\textwidth]{time_0p5u.eps}}\\ \vspace*{0.4cm}
\centering
  {\label{fig:0.6u}\includegraphics[width=0.42\textwidth]{time_0p6u.eps}} 
  {\label{fig:0.7u}\includegraphics[width=0.42\textwidth]{time_0p7u.eps}}\\ \vspace*{0.4cm}
\centering
  {\label{fig:0.8u}\includegraphics[width=0.42\textwidth]{time_0p8u.eps}} 
  {\label{fig:0.9u}\includegraphics[width=0.42\textwidth]{time_0p9u.eps}}\vspace*{0.4cm}
\caption{The time until the extinction of the slows leaving the fasts victorious as a function of $N_{0}$.}
  \label{fig:time up}
\end{figure}

In order to clarify if the apparent coexistence phase is real or only the result of the finite running time of the simulation, we measured the time until victory as the coexistence region is approached from either above (high $N_0$) or below. Figs. \ref{fig:time up} and \ref{fig:time down} display the average time until one species wins as a function of $N_{0}$. Fig. \ref{fig:time up} displays the average time until the fasts win in the small $N_0$ regime and Fig. \ref{fig:time down} displays the average time until the slows win in the higher $N_0$ regime. The time till extinction of the loser grows markedly as $N_{0}$ approaches  the coexistence area, and is consistent with a power-law divergence:
\begin{equation}
t_{f}=B_f (N_{f,c}-N_{0})^{-A_{f}}
\label{eq:time up}
\end{equation}
Likewise, the time to extinction of the losing fast species grows as $N_{0}$ approaches the coexistence area from above, again consistent with a power-law divergence:
\begin{equation}
t_{s}=B_s (N_{x}-N_{s,c})^{-A_{s}}
\label{eq:time down}
\end{equation}
The parameters of Eqs.  (\ref{eq:time up}, \ref{eq:time down}) were found by least-squares fitting. The graphs lead to the conclusion  that also for infinite time there will be no victory of one population because the winning times  diverge as the coexistence region boundary is crossed. 

\begin{figure}
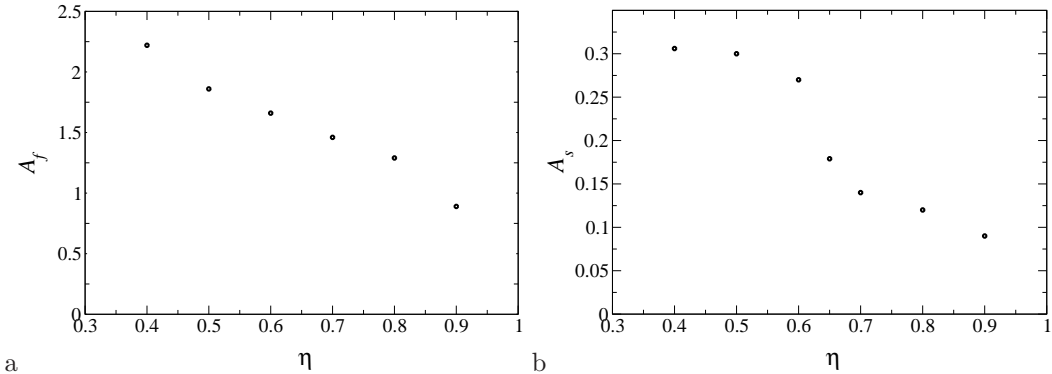

 \centering
  a{\label{fig:A1u}\includegraphics[width=0.42\textwidth]{fs_time_u.eps}}
  b{\label{fig:A1d}\includegraphics[width=0.42\textwidth]{fs_time_d.eps}}
\caption{The powers $A_{f}$, $A_s$ of the fit to the functional form Eq.(\ref{eq:time up},\ref{eq:time down}). (a) Fasts win, (b) Slows win.}
  \label{fig:A1}
\end{figure}

It is interesting to note that the exponents characterizing the divergence are different on the two sides of the transition. In addition, the exponents decrease with increasing $\eta$. Graphs of the exponents $A_{f},\ A_{s}$ as a function of $\eta$ are presented in Fig. \ref{fig:A1}.

\begin{figure}[h]
\begin{center}
 {\includegraphics[width=0.8\columnwidth]{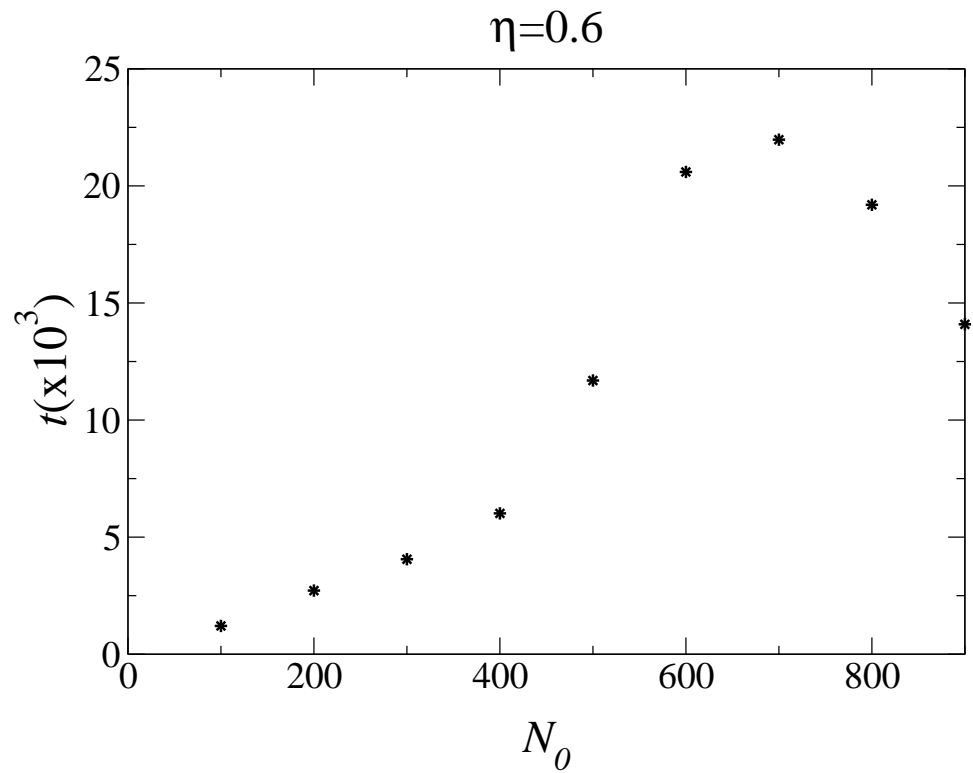}}
\end{center}
\caption{The average time until one of the population wins in the one-dimensional system, with $\eta=0.6$.}
\label{fig:time_1D}
\end{figure}

To point out the contrast between the coexistence dynamics we see in the two dimensional system and the behavior of the one dimensional simulation,
we have measured the average extinction time as a function of $N_0$ for a fixed $\eta=0.6$ in the one dimensional system.  The data is presented in Fig. \ref{fig:time_1D}. Here the extinction time rises as the border between the fast dominated and the slow dominated phases is approached, but it shows no sign of diverging.

\begin{figure}[h]
\begin{center}
\scalebox{0.9} {\includegraphics[width=\columnwidth]{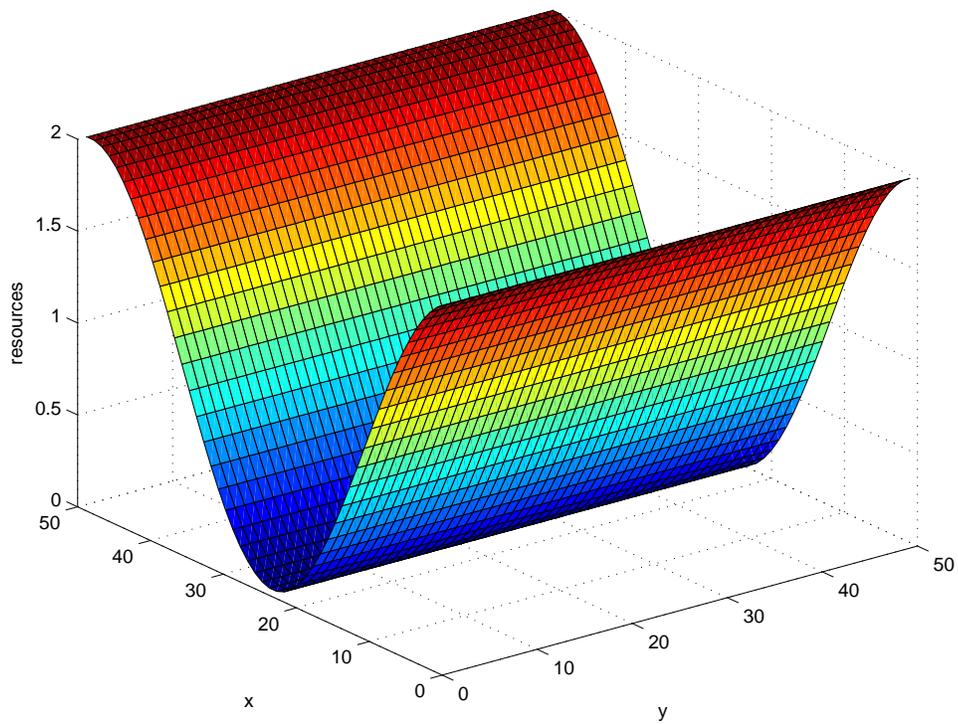}}
\end{center}
\caption{Plot of the local birthrate $\alpha(x,y)$ in the striped $50\times 50$ system.}
\label{fig:resources strip}
\end{figure}

\begin{figure}
  \centering
  a{\label{fig:fa}\includegraphics[width=0.31\textwidth]{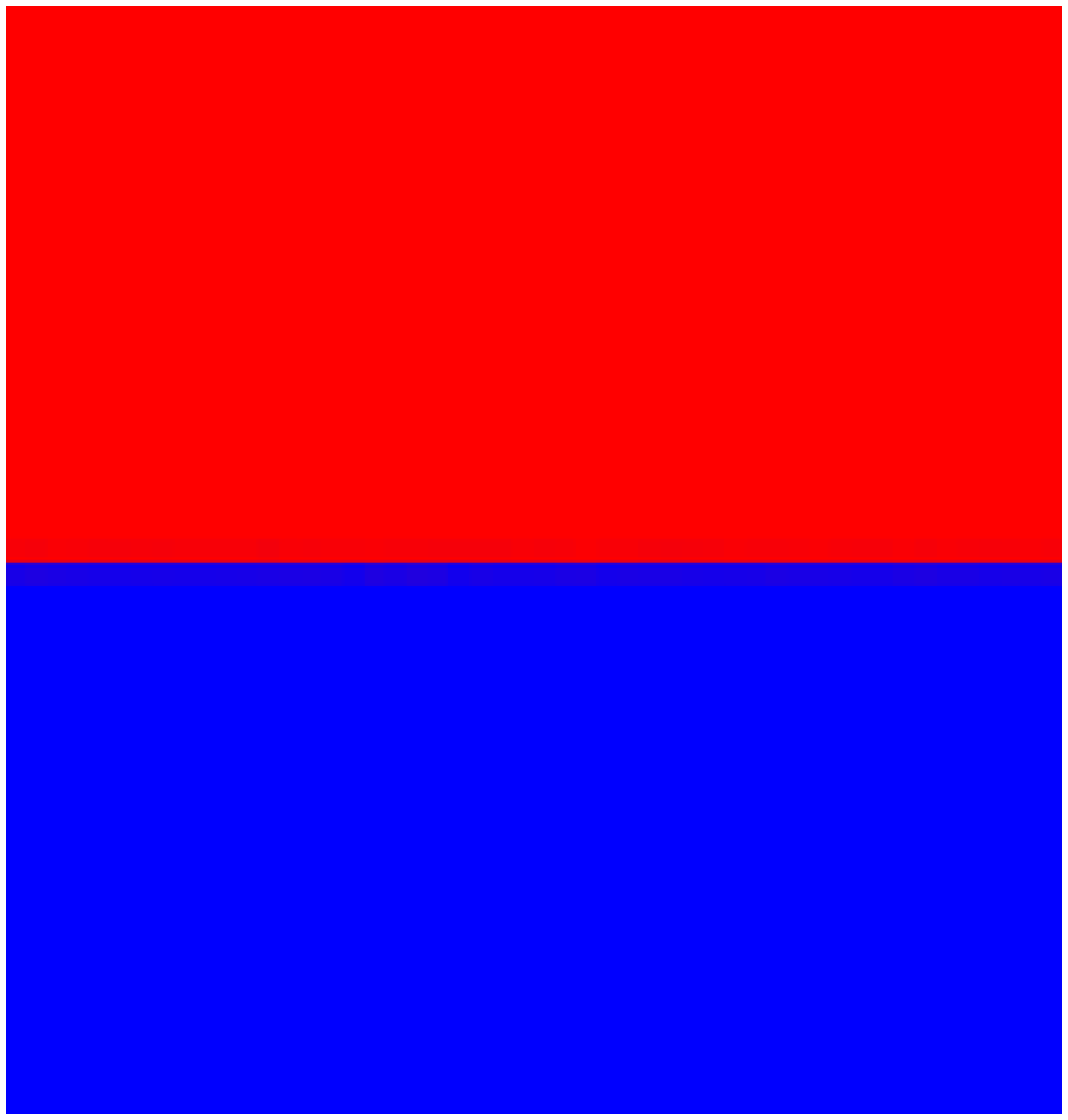}} 
  b{\label{fig:fb}\includegraphics[width=0.31\textwidth]{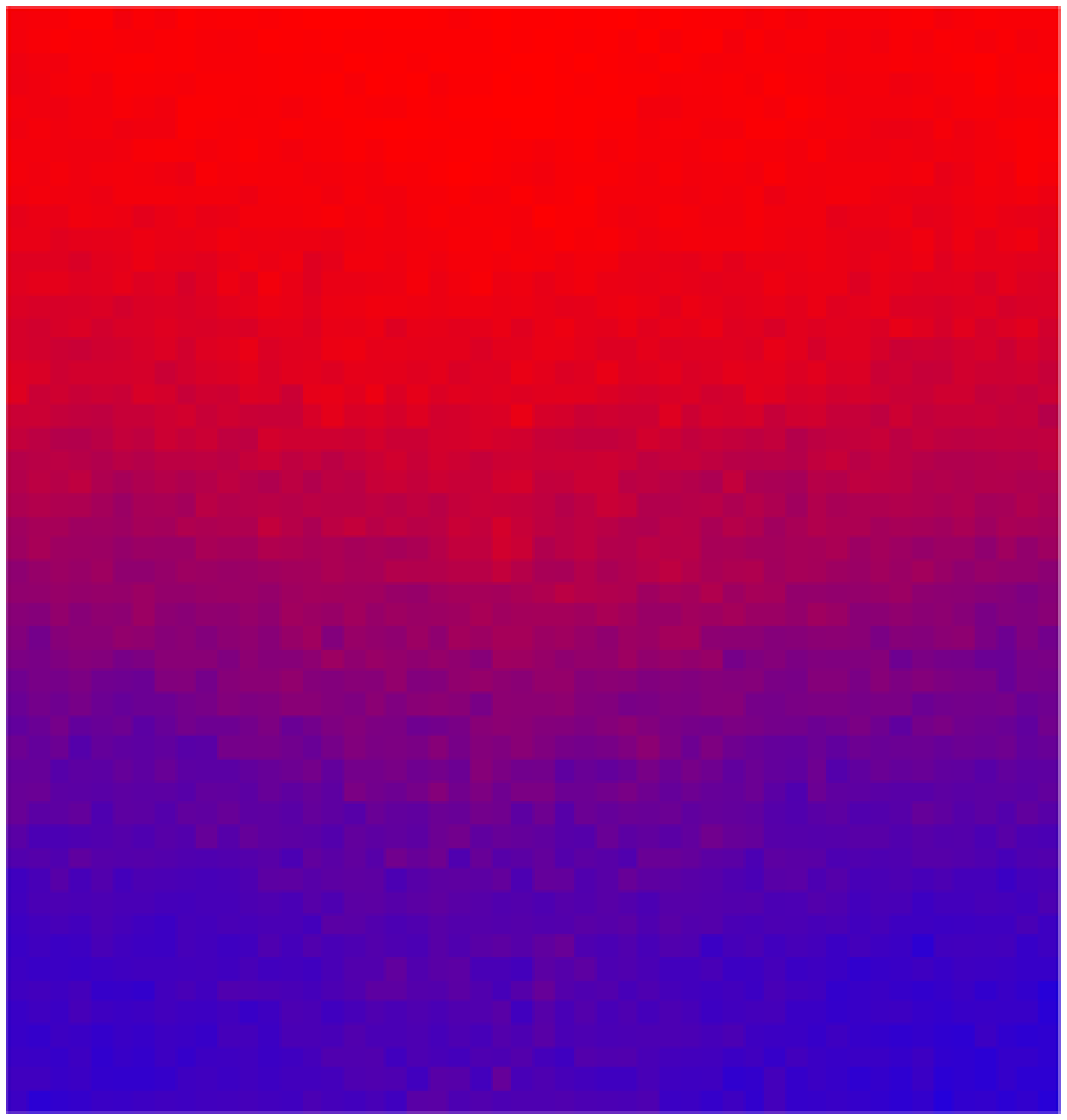}}
  c{\label{fig:fc}\includegraphics[width=0.31\textwidth]{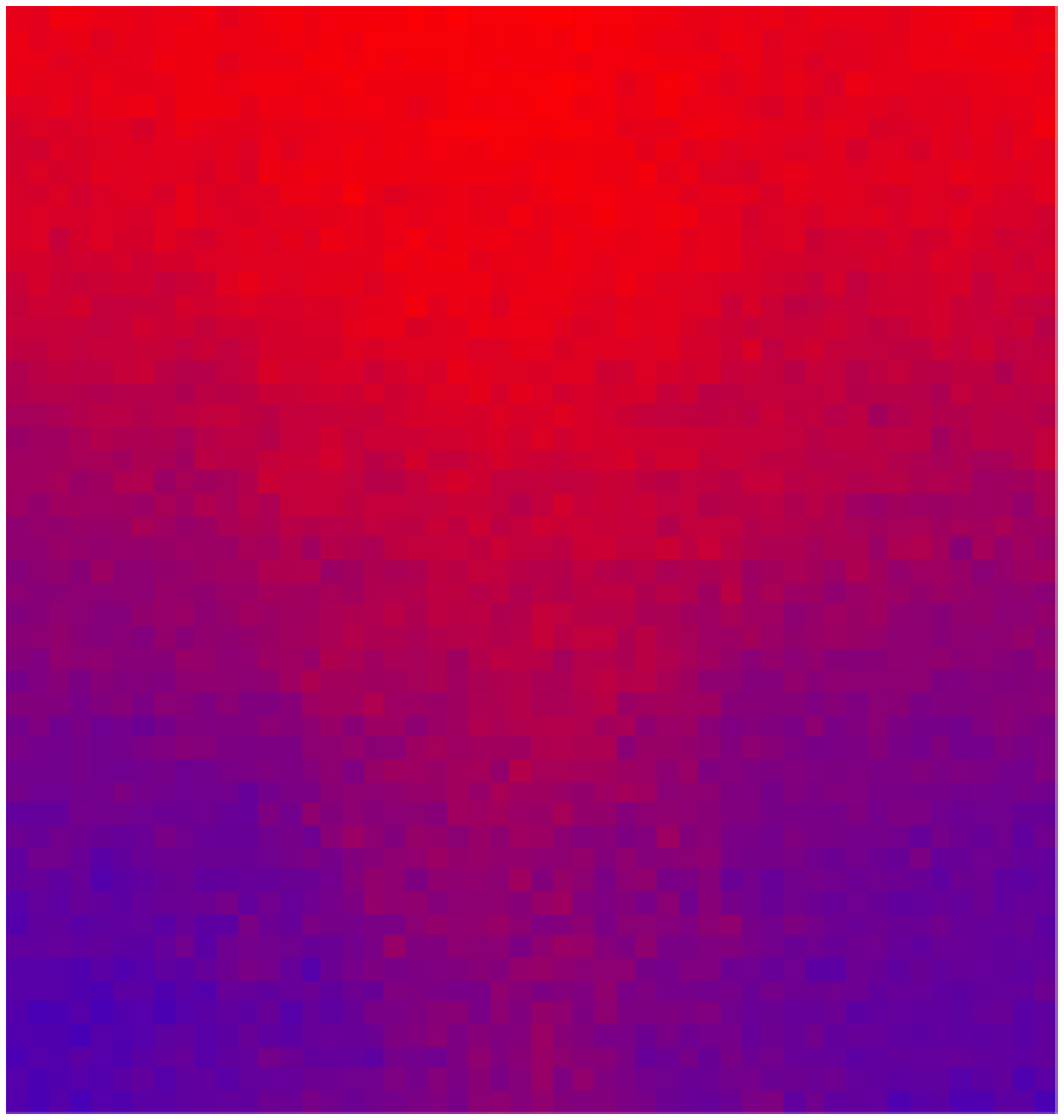}}\\
\centering
  d{\label{fig:fd}\includegraphics[width=0.31\textwidth]{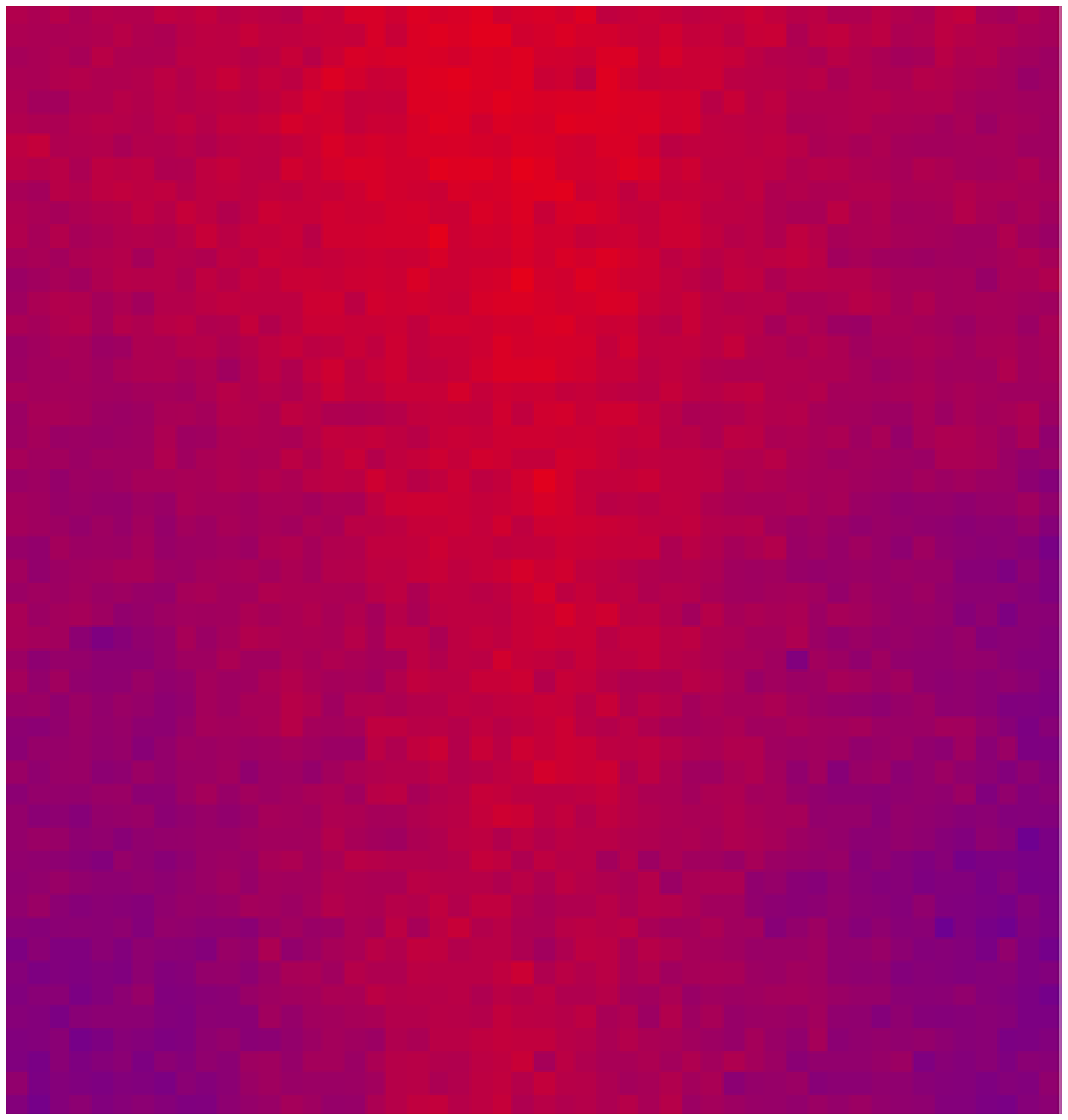}} 
  e{\label{fig:fe}\includegraphics[width=0.31\textwidth]{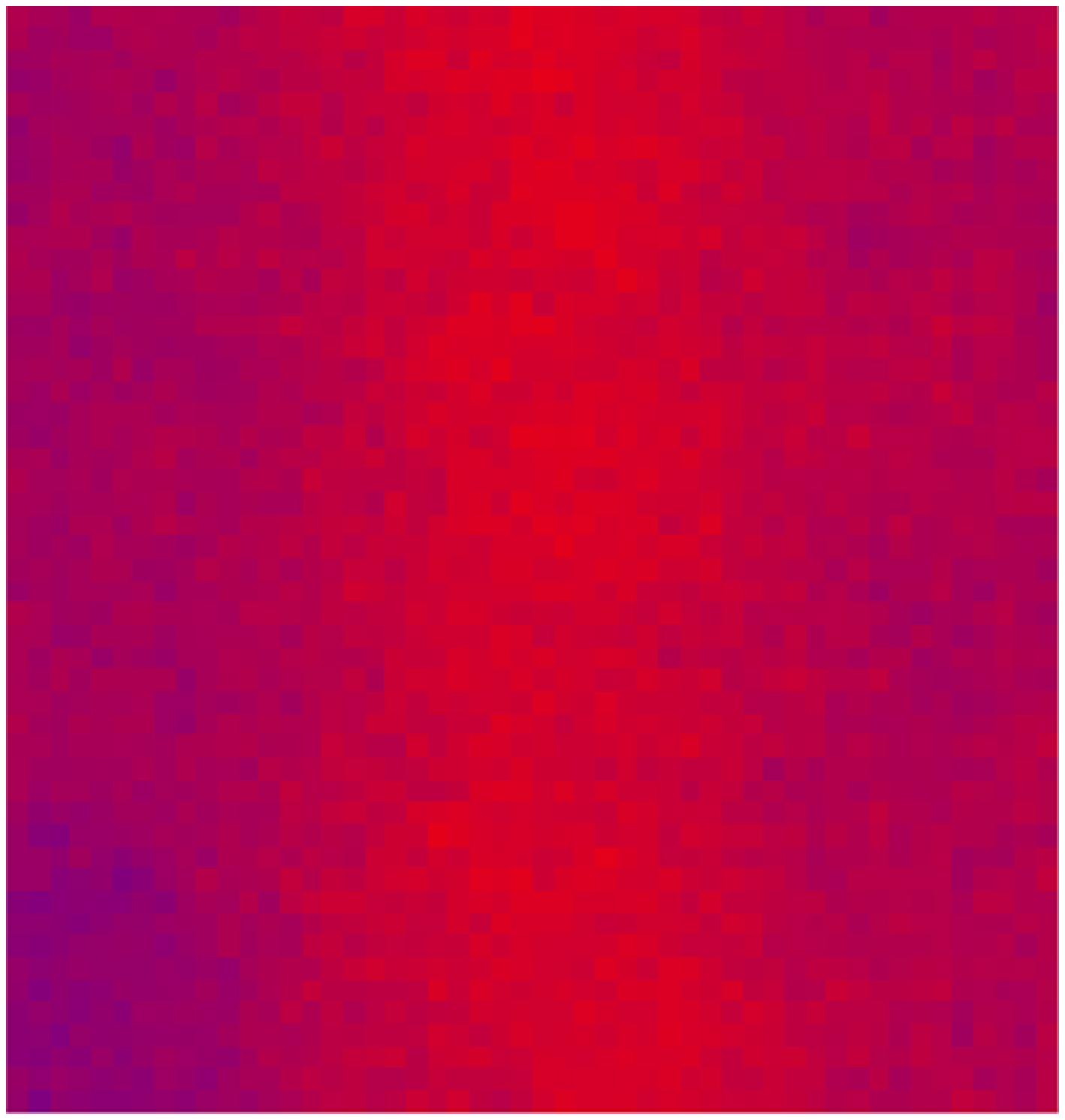}}
  f{\label{fig:ff}\includegraphics[width=0.31\textwidth]{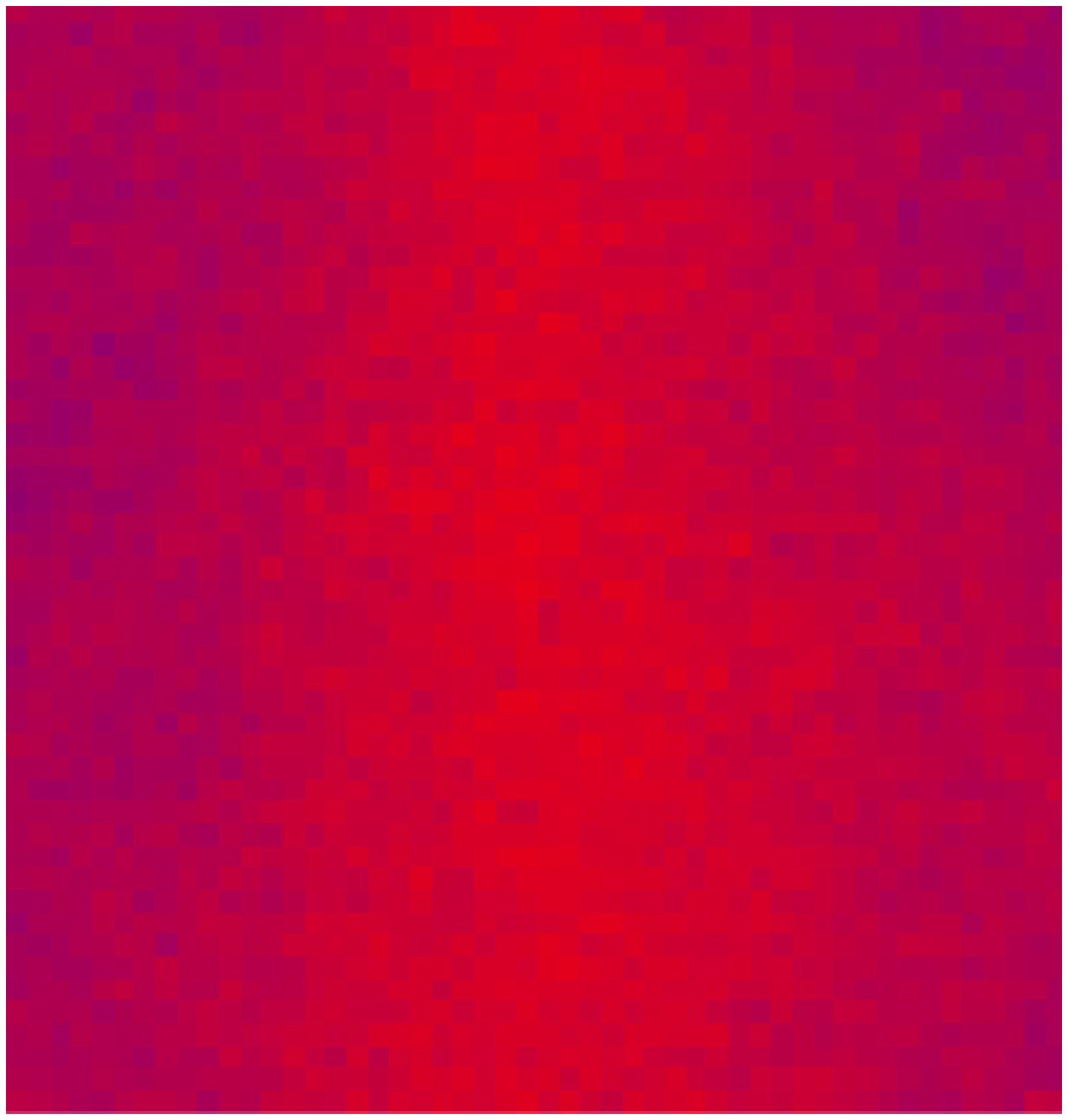}}
\caption{Graphic simulation of the striped pattern with the parameters $\eta=0.4, N_{0}=600$. (a) The initial state of the system, (b) the system after 200 steps, (c) the system after 400 steps, (d) the system after 1000 steps, (e) the system after 5000 steps, (f) the system after 20000 steps, in the (d)-(f) snapshots there are no essential changes during the time.}
  \label{fig:simstripe}
\end{figure}

\begin{figure}[h]
\begin{center}
\scalebox{0.9} {\includegraphics[width=\columnwidth]{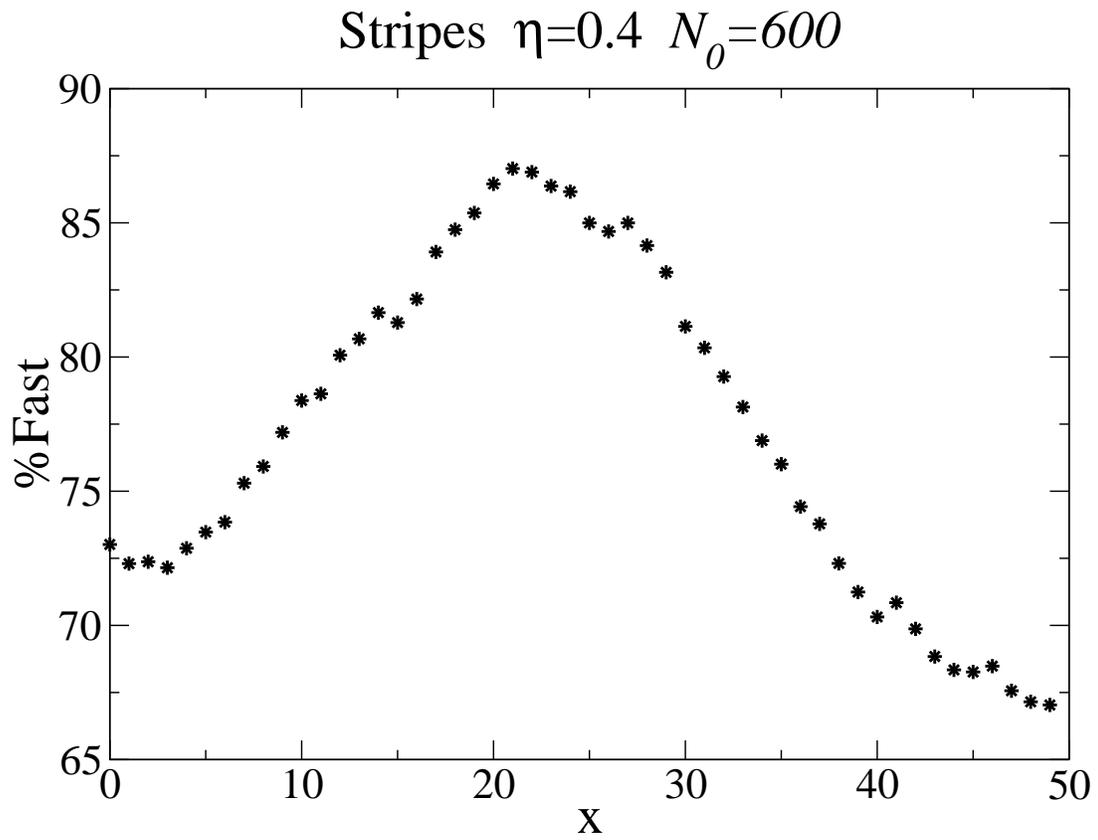}}
\end{center}
\caption{Percentange of fast individuals as a function of position in the striped two-dimensional system.  $\eta=0.4, N_{0}=600, t=20000$.}
\label{fig:perc_f_stripes}
\end{figure}

Up till now, we have focused only on one specific pattern of inhomogeneity in the two dimensional system, namely the `` checkerboard". We now investigate a ``striped" pattern of inhomogeneity, which is a uniform extension in the second dimension of the one dimensional $\alpha(x)$ (Fig. \ref{fig:resources strip}):
\begin{equation}
\alpha_\textrm{stripe}(x,y) = 1 + \eta \cos \left( \frac{2\pi(x+0.5)}{L}\right)
\end{equation}
We present snapshots from the simulation in Fig. \ref{fig:simstripe} for $\eta=0.4$, $N_0=600$, which showed coexistence in the checkerboard geometry. As always, initially the fasts dominate the ``deserts" and the slows the ``oases", but as in the checkerboard geometry, the system settles into a coexistence state. There is still a correlation between the local growth rate and the fraction of fasts in this coexistence state, but as we see in Fig. \ref{fig:perc_f_stripes},
it is surprisingly weak, with only  a $\sim 15\%$ variation between the fast/slow balance between the deserts and  the oases. 

\begin{figure}[h]
\begin{center}
\scalebox{0.9} {\includegraphics[width=\columnwidth]{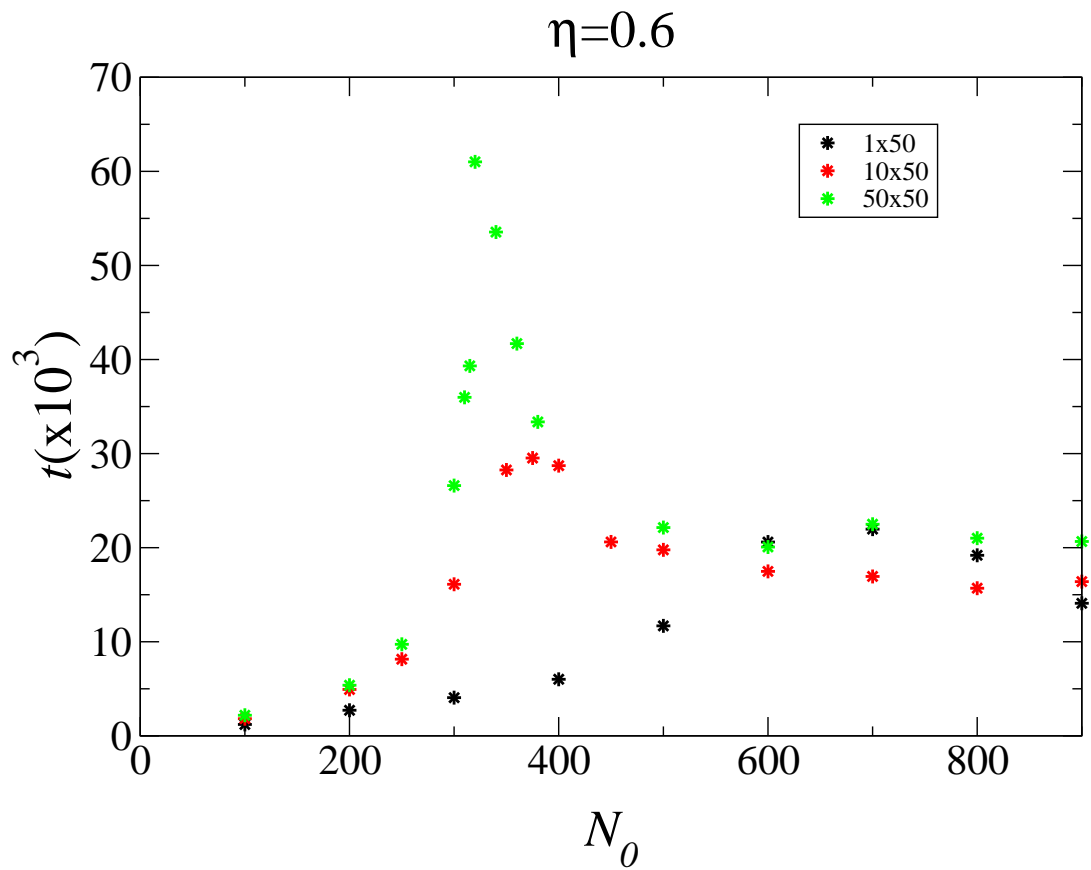}}
\end{center}
\caption{The average time until one of the population wins in the striped $1\times 50$,$10\times 50$ and $50\times 50$ system, with $\eta=0.6$.}
\label{fig:time_stripes}
\end{figure}

 Given that coexistence behavior is not present in the one-dimensional, i.e. $50\times 1$, system, but does obtain in the two-dimensional, i.e. $50\times 50$, system, the critical parameter underlying coexistence must be the width.  We show in Fig. \ref{fig:time_stripes} the average time to victory for systems of width 1, 10 and 50.  We see that as the width increases, the peak moves toward lower $N_0$ and increases in height.  The maximum time to victory for the $50\times 10 $ system is 29000 steps, which is much larger than the $50\times 1$ system, but nevertheless is finite.  We conjecture that this should be the case for any finite width system, with the maximum time to victory diverging with the system size. For the $50\times 50$ system, such a maximum time is evidently much larger than we can afford to measure.

\begin{figure}
 \centering
  {\label{fig:0.4d}\includegraphics[width=0.42\textwidth]{time_0p4d.eps}}
  {\label{fig:0.5d}\includegraphics[width=0.42\textwidth]{time_0p5d.eps}}\\ \vspace*{0.4cm}
\centering
  {\label{fig:0.6d}\includegraphics[width=0.42\textwidth]{time_0p6d.eps}} 
  {\label{fig:0.7d}\includegraphics[width=0.42\textwidth]{time_0p7d.eps}}\\ \vspace*{0.4cm}
\centering
  {\label{fig:0.8d}\includegraphics[width=0.42\textwidth]{time_0p8d.eps}} 
  {\label{fig:0.9d}\includegraphics[width=0.42\textwidth]{time_0p9d.eps}}
\caption{The time until the extinction of the fasts leaving the slows victorious as a function of $N_{0}$.}
  \label{fig:time down}
\end{figure}

\section*{Summary and Discussion}
In conclusion, we have shown that a checkerboard pattern of inhomogeneity leads to the coexistence of two different dispersal strategies, for some intermediate range of densities. This feature is absent in both one-dimensional systems and in fully-connected systems. The transition to coexistence is marked by the power-law divergence of the extinction time of the under-performing strategy. More study is required to clarify the 
nature of this coexistence phase, and the critical behavior associated with the transition.

The phenomenon of coexistence in an inhomogeneous environment is not surprising from the perspective of niche theory. The ``oases", with their high reproduction rate, are evidently favorable for the slows and the ``deserts" for the fast dispersers. However, given the density-dependent nature of the effective fitness, this is a simplistic view. In one dimension, one species may exhibit a temporary advantage, however in the fullness of time, one species always drives its competitor to extinction.  Evidently, however, in the two dimensional  environment, this is no longer true near the region where the advantage changes between species in a one dimensional patterned environment. The question of coexistence evidently requires a more refined understanding.  

Even with a standard density-independent fitness landscape, the interplay of dispersal and niches raises interesting questions worthy of further study. Dispersal will at a minimum tend to wash out the niche boundaries, with some degree of mixing at the niche boundaries. The question that arises is whether there is some degree of dispersal that wipes out completely the underlying niche structure of the environment and leads to domination by a single ``compromise" strategy.  In our system, the fairly small difference between the population balance in the ``oases'' and the ``deserts" means that dispersal effects are strong enough to wash out most, but not all, of the differences between the niches.  Given the absence of a clear boundary between the two niches, it is surprising that in the two dimensional system, coexistence is still possible. Comparing the present model to one with density-independent fitness differences should be very instructive.

\section*{Acknowledgments}
The authors thank N. Shnerb for many useful discussions.

\bibliography{nature}
%
%
%
%
%
%
%
%


\end{document}